\documentclass[sn-mathphys,Numbered]{sn-jnl}

\usepackage{braket}
\usepackage{graphicx}%
\usepackage{multirow}%
\usepackage{amsmath,amssymb,amsfonts}%
\usepackage{amsthm}%
\usepackage{mathrsfs}%
\usepackage[title]{appendix}%
\usepackage{xcolor}%
\usepackage{textcomp}%
\usepackage{manyfoot}%
\usepackage{booktabs}%
\usepackage{algorithm}%
\usepackage{algorithmicx}%
\usepackage{algpseudocode}%
\usepackage{listings}%

\usepackage[version=4]{mhchem}
\usepackage{siunitx}
\usepackage[nolist]{acronym}

\DeclareSIUnit\mob{\cm\squared\per\volt\per\second}

\theoremstyle{thmstyleone}%
\theoremstyle{thmstyletwo}%

\theoremstyle{thmstylethree}%

\begin{document}

\title[Article Title]{Upper efficiency limit of \ce{Sb2Se3} solar cells}

\author[1]{\fnm{Xinwei} \sur{Wang}}

\author[1]{\fnm{Seán R.} \sur{Kavanagh}}

\author[2]{\fnm{David O.} \sur{Scanlon}}

\author*[1,3]{\fnm{Aron} \sur{Walsh}\email{a.walsh@imperial.ac.uk}}

\affil*[1]{\orgdiv{Thomas Young Centre and Department of Materials}, \orgname{Imperial College London}, \orgaddress{\street{Exhibition Road}, \city{London}, \postcode{SW7 2AZ}, \country{UK}}}

\affil[2]{\orgdiv{School of Chemistry}, \orgname{University of Birmingham}, \orgaddress{\city{Birmingham}, \postcode{B15 2TT}, \country{UK}}}

\affil[3]{\orgdiv{Department of Physics}, \orgname{Ewha Womans University}, \orgaddress{\street{52 Ewhayeodae-gil, Seodaemun-gu}, \city{Seoul}, \postcode{03760}, \country{South Korea}}}

\begin{acronym}
\acro{1D}{one-dimensional}
\acro{TLs}{transition levels}
\acro{TL}{transition level}
\acro{CC}{configuration coordinate}
\acro{PESs}{potential energy surfaces}
\acro{CBM}{conduction band minimum}
\acro{VBM}{valence band maximum}
\acro{PL}{photoluminescence}
\acro{TAS}{thermal admittance spectroscopy}
\acro{DLTS}{deep-level transient spectroscopy}
\acro{ODLTS}{optical deep-level transient spectroscopy}
\acro{CB}{conduction band}
\acro{VB}{valence band}
\acro{VASP}{Vienna Ab initio Simulation Package}
\acro{DFT}{density functional theory}
\acro{PAW}{projector augmented-wave}
\acro{DOS}{density of states}
\acro{MPE}{multiphonon emission}
\acro{SRH}{Shockley-Read-Hall}
\acro{PV}{photovoltaic}
\end{acronym}


\abstract{
Antimony selenide (\ce{Sb2Se3}) is at the forefront of an emerging class of sustainable photovoltaic materials. Despite notable developments over the past decade, the light-to-electricity conversion efficiency of \ce{Sb2Se3} has reached a plateau of \SI{\sim 10}{\percent}. Is this an intrinsic limitation of the material or is there scope to rival the success of metal halide perovskite solar cells?  Here we assess the trap-limited conversion efficiency of \ce{Sb2Se3}. First-principles defect analysis of the hole and electron capture rates for point defects demonstrates the critical role of vacancies as active recombination centres. We predict an upper limit of \SI{25}{\percent} efficiency in \ce{Sb2Se3} grown under optimal equilibrium conditions where the concentrations of charged vacancies are minimised. We further reveal how the detrimental effect of Se vacancies can be reduced by extrinsic oxygen passivation, highlighting a pathway to achieve high-performance metal selenide solar cells close to the thermodynamic limit.

~\\
\begin{center}
\textbf{Context and Scale}
\end{center}

When exposed to light, excess charge carriers in a solar cell should be collected at the electrical contacts to generate a photocurrent. Any failure in this collection process results in energy losses. The practical performance of solar cells is significantly influenced by trap-mediated electron-hole recombination. However, it is often unclear which defects serve as the active sites, or `killer centres', for electron-hole recombination. We showcase an approach that considers the range of defects that can form in a material and predicts their abundance, trap levels, capture cross-sections, and ultimately the non-radiative recombination rates. Application to \ce{Sb2Se3} predicts that higher efficiencies are possible than comparable thin-film photovoltaic absorbers.
}


\maketitle

\section*{Introduction}
Antimony selenide (\ce{Sb2Se3}) has attracted interest as an earth-abundant and environmental-friendly alternative among thin-film photovoltaic light absorbers, owing to its suitable electronic and optical properties\cite{mavlonov2020review}. 
\ce{Sb2Se3} solar cells have achieved considerable progress since they were first reported in 2013\cite{choi2014sb2se3}, with a record conversion efficiency of \SI{10.57}{\percent}\cite{zhao2022regulating}. Nevertheless, the achieved efficiency falls far below the detailed-balance limit of \SI{\sim 30}{\percent} and lags behind the performance of other established commercial solar cells. 

Similar to other emerging photovoltaic compounds, the conversion efficiency of \ce{Sb2Se3} is primarily limited by the large open-circuit voltage (\textit{V}$_\mathrm{OC}$) deficit\cite{chen2020open}. Based on the detailed balance principle\cite{shockley1961detailed}, the unavoidable \textit{V}$_\mathrm{OC}$ deficit at \SI{300}{\kelvin} due to radiative recombination is \SIrange{0.2}{0.3}{\volt}\cite{nayak2019photovoltaic}.
A wide range of architectures (sensitised-\cite{choi2014sb2se3}/planar-\cite{zhou2014solution}type; substrate\cite{duan2022sb2se3}/superstrate\cite{zhao2022regulating} configurations), fabrication methods (rapid thermal evaporation (RTE)\cite{zhou2015thin}, close space sublimation (CSS)\cite{li2018stable}, vapor transport deposition (VTD)\cite{wen2018vapor}, chemical bath deposition (CBD)\cite{zhao2022regulating} etc.) and special treatments (post-selenisation\cite{leng2014selenization,rijal2021influence} and air exposure\cite{fleck2020oxygen,ren2023study}) have been attempted to improve the quality of \ce{Sb2Se3} devices.
However, \textit{V}$_\mathrm{OC}$ improvements have remained sluggish, with a \textit{V}$_\mathrm{OC}$ deficit $\textgreater$\SI{0.7}{\volt} for the highest-efficiency \ce{Sb2Se3} solar cell\cite{zhao2022regulating}. 

The origin of the \textit{V}$_\mathrm{OC}$ bottleneck remains under debate. One potential cause is the considerable trap density in \ce{Sb2Se3}. Defects in the absorber material reduce device performance through trap-assisted carrier recombination (Shockley-Read-Hall (SRH) recombination). 
Understanding the nature of the active defects is necessary to design strategies to minimise their impact. 
Defect characterisation techniques, such as steady-state \ac{PL} emission, \ac{TAS}, \ac{DLTS} and \ac{ODLTS}, can offer insights into trap levels, trap density and defect capture cross-sections.
The identification of the defect type, however, is often difficult for experiments and relies heavily on theoretical results. 
Point defects in \ce{Sb2Se3} have been widely studied by first-principles calculations\cite{liu2017enhanced,savory2019complex,huang2019complicated,stoliaroff2020deciphering,huang2021more,wang2023four}, where thermodynamic transition levels were predicted. 
The community has tried to identify the most detrimental defect in \ce{Sb2Se3} by matching measured defect levels with theoretical results.
Nevertheless, due to the complexity of defect physics of \ce{Sb2Se3}, there has been a debate on whether antisites or vacancies are the most detrimental `killer' imperfections\cite{zeng2016antimony,tang2019highly,duan2022sb2se3,zhao2022regulating}.
On the other hand, defects with deep levels were proposed as potential recombination centres, whereas the depth alone is not a sufficient condition for rapid electron and hole capture processes\cite{dou2023chemical,kavanagh2021rapid}.
Moreover, recent work has shown that global optimisation of defect geometries is important to obtain the true ground-state structures and behaviour (e.g. energy levels and recombination activity),\cite{wang2023four,mosquera2021search} with this being particularly important in lower-symmetry materials, calling into question conclusions based on singular defect relaxations.

In this work, we have investigated the intrinsic point defects in \ce{Sb2Se3} using a global structure searching strategy\cite{mosquera2022shakenbreak,mosquera2023identifying,wang2023four}, and have studied the non-radiative carrier capture processes by systematic first-principles calculations. 
The upper limit to the conversion efficiency in \ce{Sb2Se3} is predicted by considering both radiative and non-radiative processes, acting as a quantitative measure of defect tolerance. 
Vacancies are identified as the most detrimental recombination centres, with the largest contributions coming from \textit{V}$_\mathrm{Se}$ and \textit{V}$_\mathrm{Sb}$ under Se-poor and Sb-poor conditions, respectively. 
We conclude that \ce{Sb2Se3} solar cells suffer from significant non-radiative recombination, especially under extreme Sb-rich growth conditions, and higher conversion efficiencies can be achieved under intermediate growth conditions which minimise vacancy concentrations. 
The impact of oxygen passivation is further studied, demonstrating its effectiveness in enhancing the performance of \ce{Sb2Se3} by transforming the deep levels associated with detrimental Se vacancies to shallow ones.
These results elucidate the loss mechanisms associated with intrinsic point defects and provide insights into optimising the performance of \ce{Sb2Se3} solar cells.


\section*{Results and Discussion}

\subsection*{Equilibrium point defect population}

\begin{figure}[h!]
    \centering    {\includegraphics[width=0.5\textwidth]{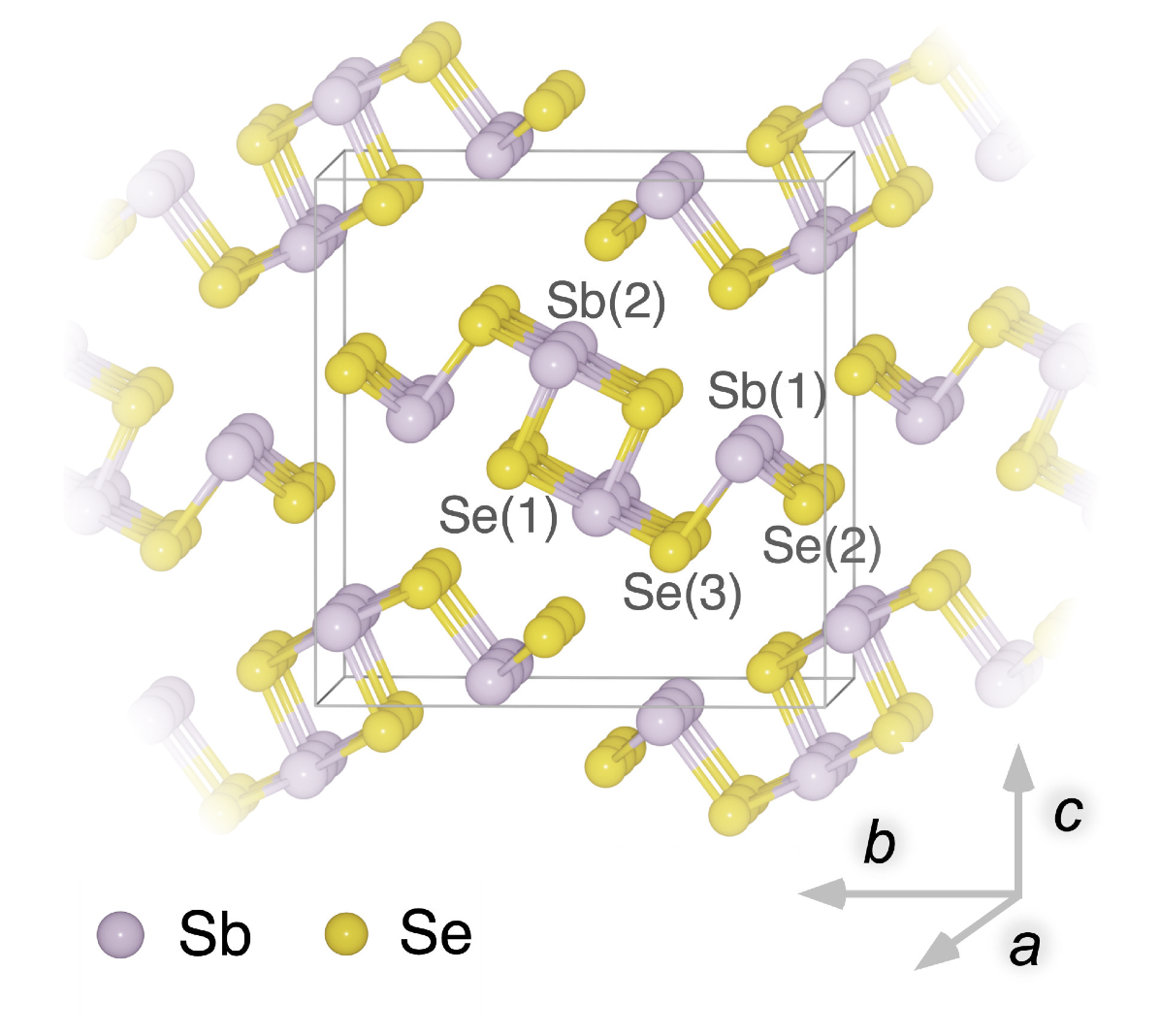}} \\
    \caption{Ground-state crystal structure (\textit{Pnma} space group) of \ce{Sb2Se3}. The unit cell is represented by a cuboid. Inequivalent sites are denoted by the atom labels enclosed in parentheses.}
    \label{fig_structure}
\end{figure}

\ce{Sb2Se3} adopts an orthorhombic crystal structure, Fig. \ref{fig_structure}. 
The structure is composed of quasi-\ac{1D} [Sb$_4$X$_6$]$_n$ ribbons arranged together via weak interactions\cite{wang2022lone}. Due to the low crystal symmetry, the chemical environment for each Sb/Se element in the unit cell is different, leading to two inequivalent Sb sites and three inequivalent Se sites. 

\textbf{\begin{figure}[h!]
    \centering    {\includegraphics[width=0.9\textwidth]{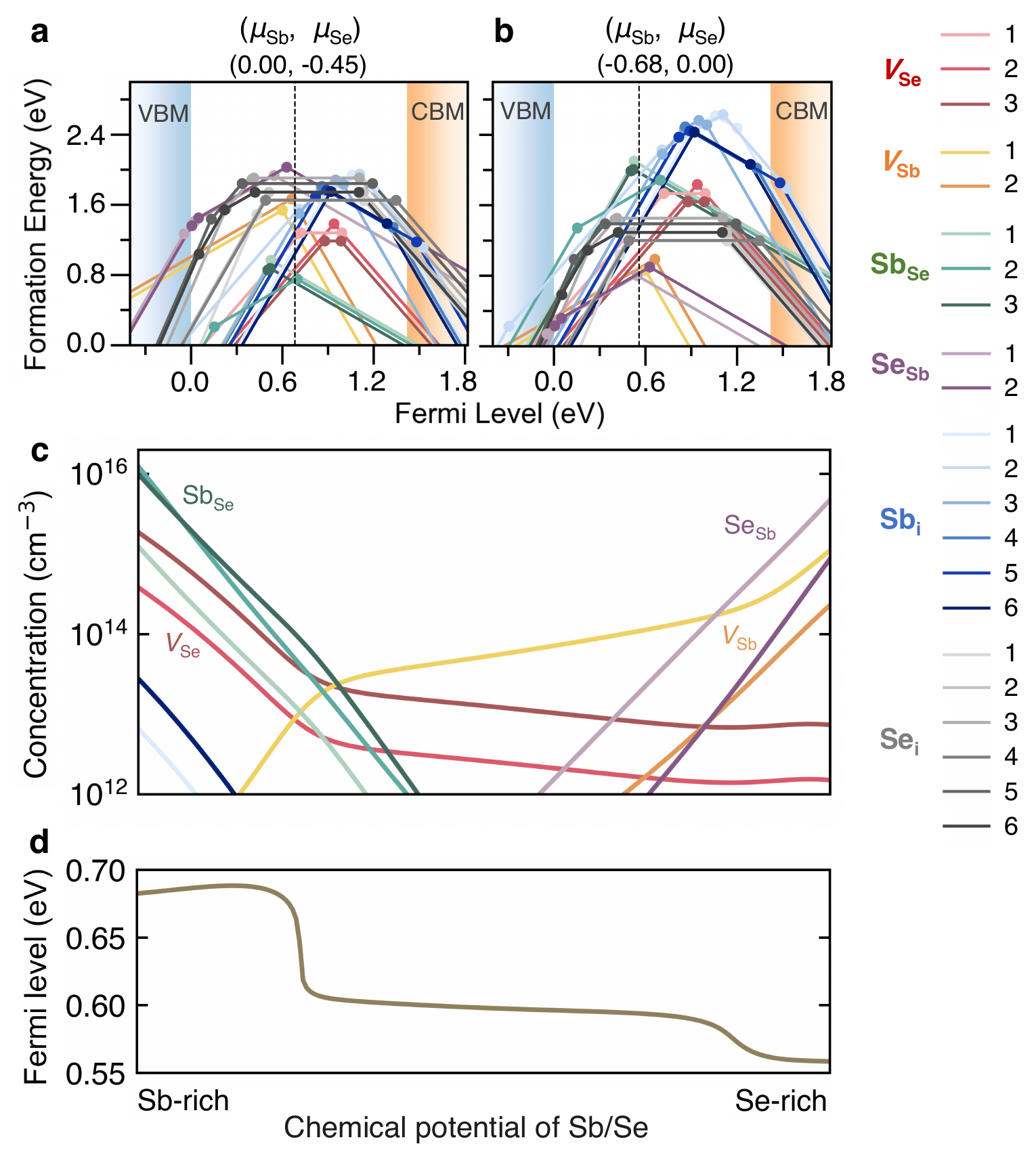}} \\
    \caption{(a)-(b) Calculated formation energies of intrinsic point defects in \ce{Sb2Se3} under chemical potentials that represent (a) Sb-rich and (b) Se-rich growth conditions. The charge state is denoted by the slope of the solid line, and the thermodynamic transition level corresponds to the filled circle. The valence band maximum (VBM) is set to \SI{0}{eV}, and the conduction band minimum (CBM) is obtained from the calculated fundamental (indirect) band gap of \SI{1.42}{eV} by the HSE06 functional. The self-consistent Fermi levels are indicated by vertical dashed lines. The numbers in the legend represent different inequivalent sites. (c) Equilibrium defect concentration and (d) self-consistent Fermi level (relative to the VBM) at \SI{300}{\kelvin} in \ce{Sb2Se3} crystals grown at \SI{550}{\kelvin}\cite{huang2021more,zhao2022regulating,duan2022sb2se3} as a function of the growth condition.}
    \label{fig_conc}
\end{figure}}

We first investigate all intrinsic point defects (i.e. vacancies, antisites and interstitials) in \ce{Sb2Se3}.
The \textsc{ShakeNBreak}\cite{mosquera2022shakenbreak,mosquera2023identifying} structure-searching workflow is applied for each defect species to identify the ground-state geometries.
All inequivalent sites are considered, giving rise to five types of vacancies (\textit{V}$\mathrm{_{Sb(1)}}$, \textit{V}$\mathrm{_{Sb(2)}}$,  \textit{V}$\mathrm{_{Se(1)}}$, \textit{V}$\mathrm{_{Se(2)}}$ and \textit{V}$\mathrm{_{Se(3)}}$) and five antisites (Se$\mathrm{_{Sb(1)}}$, Se$\mathrm{_{Sb(2)}}$,  Sb$\mathrm{_{Se(1)}}$, Sb$\mathrm{_{Se(2)}}$ and Sb$\mathrm{_{Se(3)}}$). Besides, nine inequivalent sites for interstitials Sb$\mathrm{_{i}}$/Se$\mathrm{_{i}}$ (shown in Fig. S1) are considered as initial defect configurations by the Voronoi scheme, which has been shown to be an efficient approach for sampling interstitial sites\cite{goyal2017computational,kononov2023identifying}.
Subsequent geometry relaxation yields six distinct interstitial configurations. 
The workflow of generating and optimising the defect structures is discussed in Methods.
%
Formation energies of all defects are calculated under different equilibrium growth conditions (Fig. \ref{fig_conc}(a) and (b) for Sb-/Se-rich conditions and Fig. S2 for Se-moderate conditions). 
We find that no native defects are of high energy in this system, all being in the $\textless$\SI{2.6}{eV} range, which can be partly attributed to the soft crystal structure and chemical bonding --- akin to lead halide perovskites\cite{motti2019defect}.
One unique feature of this system is that all intrinsic point defects show amphoteric behaviour, with both stable positively and negatively charged states. For defects with the lowest formation energies, all thermodynamic \ac{TLs} are very deep.
Moreover, defect behaviour can differ significantly for different inequivalent sites. For example, \textit{V}$\mathrm{_{Se(2)}}$ exhibits an unusual four-electron negative-\textit{U} behaviour (i.e. negative electron (pair) correlation energy and two thermodynamically stable charge states differing by 4 electrons; $\Delta q$ = 4), whereas \textit{V}$\mathrm{_{Se(1)}}$ and \textit{V}$\mathrm{_{Se(3)}}$ only show typical two-electron negative-\textit{U} transitions ($\Delta q$ = 2)\cite{wang2023four}.

The formation energies of defects change significantly as the growth conditions change from Sb-rich to Se-rich. The equilibrium defect concentration is further calculated as a function of the growth condition. As shown in Fig. \ref{fig_conc}(c), the dominant defects with high concentrations (\textgreater \SI{e14}{\per\cubic\cm}) under Sb-rich (Se-rich) condition are Sb$_\mathrm{Se}$ and \textit{V}$_\mathrm{Se}$ (Se$_\mathrm{Sb}$ and \textit{V}$_\mathrm{Sb}$), while the concentrations of all interstitials are low despite the open crystal structure.
The antisites/vacancies benefit from energy lowering reconstructions (valence alternation)\cite{wang2023four} that increase their concentrations.
Heavy charge compensation from the amphoteric defects results in low carrier concentrations in the dark of around \SI{e8}{\per\cubic\cm} and \SI{e10}{\per\cubic\cm} under Sb-rich and Se-rich conditions respectively (Fig. S3), which qualitatively match experimental observations of low carrier concentrations in \ce{Sb2Se3}\cite{mavlonov2020review}. 

The self-consistent Fermi level (E$_F$) as a function of the growth condition is shown in Fig. \ref{fig_conc}(d), which is pinned close to the middle of the band gap owing to strong charge compensation from the low energy defects. With the increase of $\mu_\mathrm{Se}$, the self-consistent E$_F$ decreases from \SI{0.68}{eV} under Sb-rich conditions to \SI{0.56}{eV} under Se-rich conditions, which are in good agreement with the experimental results of \SI{0.60}{eV} and \SI{0.52}{eV} under Se-poor and Se-rich conditions, respectively\cite{ma2019fabrication}.
Considering the calculated fundamental band gap of \SI{1.42}{eV}, this indicates intrinsic weakly \textit{p}-type conductivity, which agrees well with the naturally weak \textit{p}-type behaviour in \ce{Sb2Se3} reported by most studies\cite{liu2014thermal,zhou2014solution,zhou2017buried}.

It is worth noting that Huang et al.\cite{huang2021more} predicted a higher concentration of \textit{V}$_\mathrm{Se}$ under Se-rich compared to Se-poor conditions by first-principles calculations, an unusual situation driven by the Fermi level changes. 
The main origin of this difference is our identification of low-energy positive charge states for \textit{V}$_\mathrm{Se}$, \textit{V}$_\mathrm{Sb}$ and Se$_\mathrm{Sb}$ under Se-rich conditions using a global structure searching strategy\cite{mosquera2022shakenbreak,mosquera2023identifying,wang2023four}. This results in strong charge compensation and a self-consistent Fermi level near midgap (\SI{0.56}{eV}), which matches well with the experimental value of \SI{0.60}{eV}\cite{ma2019fabrication}. 
Our predicted intrinsic midgap Fermi level corresponds to higher formation energies for \textit{V}$^{2+}_\mathrm{Se}$, and thus much lower predicted \textit{V}$_\mathrm{Se}$ concentrations under Se-rich conditions.
We therefore find that global structure searching is necessary to accurately predict defect properties in chalcogenide semiconductors.

\subsection*{Non-equilibrium carrier capture}

\begin{figure}[h]
    \centering    {\includegraphics[width=\textwidth]{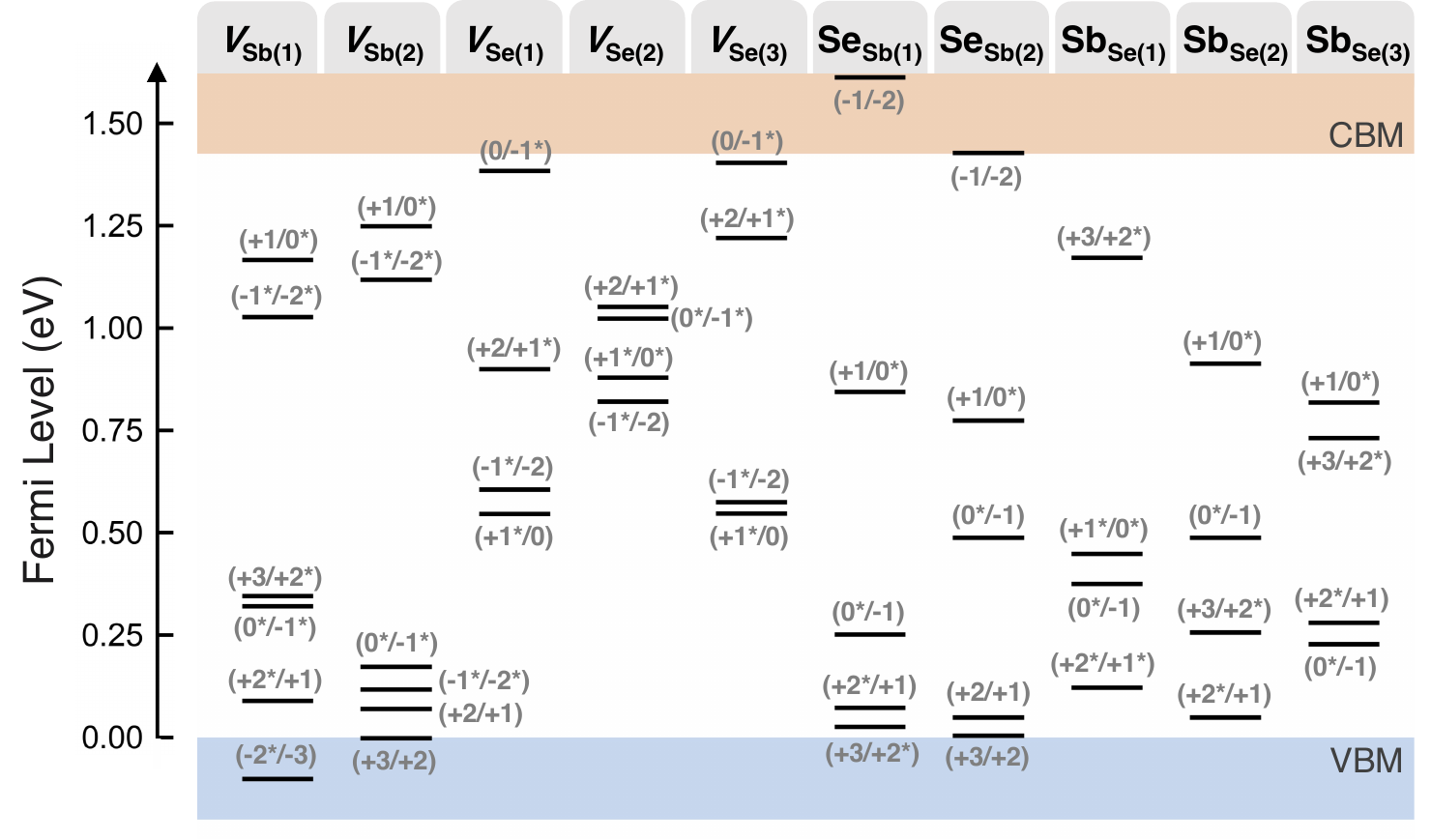}} \\
    \caption{Charge state transition levels of intrinsic point defects with high concentrations in \ce{Sb2Se3}. Metastable charge states (which are not the thermodynamic ground-state at any Fermi level position; Fig. \ref{fig_conc}) are indicated with asterisks (*), and the Fermi level is referenced to the valence band maximum (VBM). }
    \label{fig_tl}
\end{figure}


We next consider the kinetics of non-radiative carrier capture and recombination when \ce{Sb2Se3} is subject to above bandgap illumination.
A description of the microscopic processes requires going beyond the static defect properties and consideration of the dynamics of transitions between different charge states of a defect. 
This is achieved by introducing \ac{CC} diagrams that connect the initial (charge \textit{q}) and final (charge \textit{q')} state structures of each defect D. Trap-mediated electron-hole recombination can be considered in terms of the successive capture of electrons and holes, i.e.
 \begin{equation}
 \ce{
 D^q <=>[h\nu]
 D^q + e^- + h^+ <=>[-\hbar\omega] D^{q'} + h^+ <=>[-\hbar\omega] D^q}
 \end{equation}
where the excess electronic energy provided by light absorption ($h\nu$) is thermally emitted through phonons ($\hbar\omega$).

We start by considering the single-electron transitions for those defects with high concentrations (i.e. all vacancies and antisites). 
The single-electron transition energy levels are shown in Fig. \ref{fig_tl}. Multiple inequivalent sites and accessible charge states make the defect levels in \ce{Sb2Se3} complex to analyse. Different transition levels share similar energy ranges as shown in Fig. \ref{fig_tl}. Consequently, it is difficult to identify the defect species solely based on the comparison of energy levels with values that are measured experimentally.
Since our objective is to identify potential recombination centres with both rapid electron and hole capture, shallow defect levels (i.e. where the defect level and band edge energy difference is comparable to the thermal energy $k_\textrm{B}T$) are excluded from consideration.

\begin{figure}[h]
    \centering    {\includegraphics[width=\textwidth]{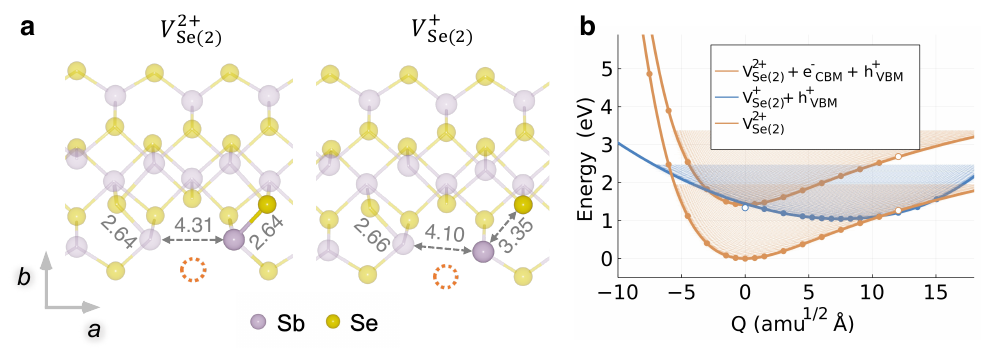}} \\
    \caption{(a) Defect configurations of \textit{V}$\mathrm{^{2+}_{Se(2)}}$ and \textit{V}$\mathrm{^{+}_{Se(2)}}$. The bond lengths in Å are labelled, and the vacant Se site is denoted by a dotted circle. (b) One-dimensional configuration coordinate diagram for charge transitions between \textit{V}$\mathrm{^{2+}_{Se(2)}}$ and \textit{V}$\mathrm{^{+}_{Se(2)}}$. Solid circles are data points obtained by DFT calculations and used for fitting, while hollow circles are discarded for fitting due to charge delocalisation (see Section S4.1.1 of SI). Solid lines represent best fits to the data.}
    \label{fig_pes}
\end{figure}

The complete pathways for trap-mediated electron and hole capture by point defects, including those introduced by low-energy metastable states (shown to be important for accurate predictions\cite{kavanagh2022impact}), are mapped (shown in Fig. S4). 
The dominant charge-capture transition under most growth conditions, having both high defect concentrations and large electron and hole capture coefficients, is predicted to be \textit{V}$^{2+}_\mathrm{Se(2)}$ $\leftrightarrow$ \textit{V}$^{+}_\mathrm{Se(2)}$. The corresponding atomic structures and \ac{PESs} are shown in Fig. \ref{fig_pes}. Structures and PESs for other charge-capture transitions can be found in Section S5 of SI. Table \ref{table_pes} shows the carrier capture coefficients and cross-sections at room temperature, and key parameters used in the calculations. 

\begin{table*}[ht!]
\caption{Key parameters used to calculate the carrier capture coefficients in the transition of \textit{V}$^{2+}_\mathrm{Se(2)}$ $\leftrightarrow$ \textit{V}$^{+}_\mathrm{Se(2)}$: mass-weighted distortion $\Delta$\textit{Q} (amu$^{1/2}$\AA), energy barrier $\Delta$\textit{E}$_\textrm{b}$  (meV), degeneracy factor \textit{g} of the final state, electron-phonon coupling matrix element $W_{if}$ and scaling factor \textit{s}(\textit{T})\textit{f} at \SI{300}{\kelvin}, along with calculated capture coefficient \textit{C} (\SI{}{\cubic\cm\per\second}) and cross-section $\sigma$ (\SI{}{\cm\squared}) at \SI{300}{\kelvin}}
\label{table_pes}
\begin{tabular*}{\textwidth}{@{\extracolsep{\fill}}c@{\extracolsep{\fill}}c@{\extracolsep{\fill}}c@{\extracolsep{\fill}}c@{\extracolsep{\fill}}c@{\extracolsep{\fill}}c@{\extracolsep{\fill}}c@{\extracolsep{\fill}}c}
    \hline
$\Delta$\textit{Q} & \begin{tabular}[c]{@{}c@{}}Capture\\ process\end{tabular} & $\Delta$\textit{E}$_\textrm{b}$ & \textit{g} & $W_{if}$ & \textit{s}(\textit{T})\textit{f} & \textit{C} & $\sigma$ \\     \hline
\multirow{2}{*}{7.52} & Electron & \SI{2}{} & 4 & \SI{1.81e-2}{} & 2.09  &\SI{5.63e-6}{} & \SI{2.85e-13}{} \\
  & Hole & \SI{83}{} & 1 & \SI{1.76e-2}{} & 0.35 &\SI{1.22e-8}{} & \SI{9.89e-16}{} \\     \hline
\end{tabular*}
\end{table*}

The mass-weighted displacement $\Delta Q$ represents the structural difference between the two defect charge states involved in the charge-capture process.
The main contribution to $\Delta Q$ of 7.52 amu$^{1/2}${\AA } for \textit{V}$^{2+}_\mathrm{Se(2)}$ and \textit{V}$^{+}_\mathrm{Se(2)}$ comes from the shortening/lengthening of one Sb-Se bond length beside \textit{V}$_\mathrm{Se(2)}$ (highlighted in Fig. \ref{fig_pes}(a)) during the hole/electron capture process.
\ac{PESs} were mapped by performing single-point DFT calculations for interpolated configurations between the equilibrium structures of \textit{V}$^{2+}_\mathrm{Se(2)}$ and \textit{V}$^{+}_\mathrm{Se(2)}$ (Fig. \ref{fig_pes}(a)). 
The electronic eigenstates at each \textit{Q} were checked (Fig. S6) to remove any datapoints where the occupation of single-particle defect levels changed due to crossing the band edges (i.e. charge delocalisation) from fitting.
The equilibrium structure of \textit{V}$^{2+}_\mathrm{Se(2)}$ is set as a reference with \textit{Q} = 0 amu$^{1/2}${\AA } and \textit{E} = \SI{0}{\electronvolt}. The equilibrium structure of \textit{V}$^{+}_\mathrm{Se(2)}$ is offset horizontally by $\Delta {Q}$ and vertically by $\Delta {E}$ = \SI{1.05}{\electronvolt} (which corresponds to position of the (+2/+1) transition level with respect to the \ac{VBM}). The uppermost orange curve (\textit{V}$^{2+}_\mathrm{Se(2)} + \textrm{e}^- + \textrm{h}^+$) is vertically upshifted by the fundamental band gap $\textit{E}_g$ compared to the bottom-most orange curve (\textit{V}$^{2+}_\mathrm{Se(2)}$), corresponding to the energy of the photo-excited electron-hole pair.
Further details regarding the calculation of carrier capture coefficients via the configuration coordinate approach are given in Methods.

In the process of non-radiative capture of an electron by \textit{V}$^{2+}_\mathrm{Se(2)}$, the initial (excited) state is represented by the uppermost orange curve, and the final (ground) state corresponds to the blue curve. 
The two PESs intersect at $\Delta$\textit{E}$_\textrm{b}$ = \SI{2}{\meV} above the minimum of the excited state.
The negligible $\Delta$\textit{E}$_\textrm{b}$ and large phonon overlap result in a large electron capture coefficient ($C_p$) of \SI{5.63e-6}{\cubic\cm\per\second} at room temperature.
In the non-radiative capture of a hole by \textit{V}$^{+}_\mathrm{Se(2)}$, the initial and final states correspond to the blue and bottom-most orange curves, respectively. A weaker Coulomb repulsion of positively-charged holes by \textit{V}$^{+}_\mathrm{Se(2)}$ (included in the scaling factor \textit{s}(\textit{T})\textit{f}), reduced pathway degeneracy $g$ and larger $\Delta$\textit{E}$_\textrm{b}$ of \SI{83}{\meV} (Table \ref{table_pes}) all contribute to a smaller hole capture coefficient ($C_p$) of \SI{1.22e-8}{\cubic\cm\per\second} at room temperature.
Therefore, electron-hole carrier recombination at\textit{V}$_\mathrm{Se(2)}$ is limited by the hole capture process \textit{V}$\mathrm{^{+}_{Se(2)}} + \textrm{h}^+ \rightarrow$ \textit{V}$\mathrm{^{2+}_{Se(2)}}$. The calculated capture cross-sections ($\sigma$) agree well with the range of experimental results (10$^{-17}$--10$^{-13}$ cm$^2$)\cite{zhao2022regulating,wen2018vapor}.

\subsection*{Trap-limited conversion efficiency}

\begin{figure*}[ht!]
    \centering   {\includegraphics[width=\textwidth]{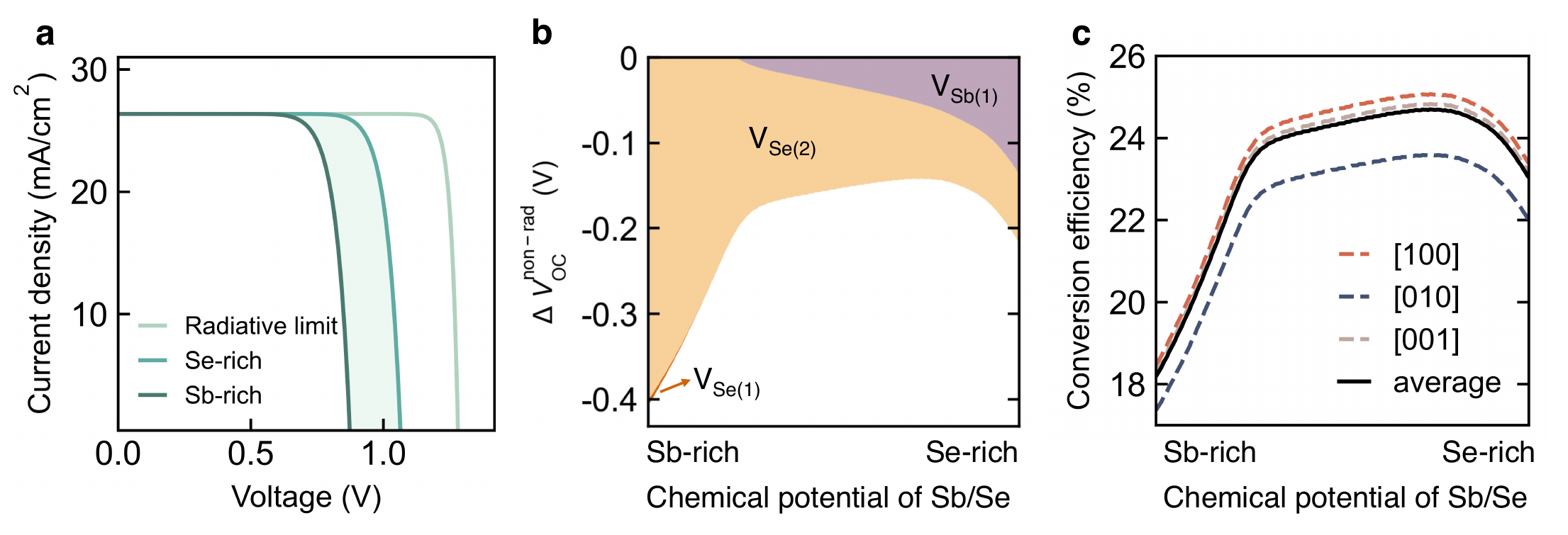}} \\
    \caption{(a) Calculated current density-voltage ($J-V$) curves for \ce{Sb2Se3}, assuming the radiative limit (only band-to-band radiative recombination losses) and including defect-induced non-radiative recombination under Se/Sb-rich growth conditions. (b)\textit{V}$\mathrm{_{OC}}$ deficit contributed by non-radiative recombination ($\Delta$\textit{V}$\mathrm{_{OC}^\mathrm{non-rad}}$) in undoped \ce{Sb2Se3} as a function of growth condition, decomposed into individual defect contributions. $\Delta$\textit{V}$\mathrm{_{OC}^\mathrm{non-rad}}$ is defined as the difference between the values of \textit{V}$\mathrm{_{OC}}$ and \textit{V}$\mathrm{_{OC}^\mathrm{rad}}$. (c) Trap-limited conversion efficiency as a function of the growth condition. [100], [010] and [001] correspond to the crystallographic directions in \ce{Sb2Se3}. All results shown correspond to a film thickness of 500 nm, and room temperature defect concentrations assuming an annealing temperature of 550 K\cite{huang2021more,zhao2022regulating,duan2022sb2se3}.}
    \label{fig_eff}
\end{figure*}

To directly quantify the impact of electron-hole recombination at point defects on the \ce{Sb2Se3} solar cell performance, the conversion efficiency is predicted using the aTLC model\cite{kim2020upper,kim2021ab}. Detailed equations can be found in Methods.
Current density-voltage ($J-V$) curves of \ce{Sb2Se3} solar cells are predicted under different growth conditions including both radiative and non-radiative recombination (Fig. \ref{fig_eff}(a)). Directionally-averaged optical absorption coefficients are used when calculating the radiative limit, considering the polycrystalline nature of most \ce{Sb2Se3} samples from the literature. The film thickness is set to \SI{500}{\nm}, which has been found to maximise short-circuit current density (\textit{J}$\mathrm{_{SC}}$) experimentally\cite{li2020simulation}. 
Open-circuit voltage \textit{V}$\mathrm{_{OC}}$ deficit (defined as $E_g/\textit{q}-\textit{V}\mathrm{_{OC}}$) due to radiative recombination is predicted to be \SI{0.14}{/V}.
Non-radiative recombination is found to significantly contribute to \textit{V}$\mathrm{_{OC}}$ deficit, with predicted total deficits of \SI{0.54}{/V} and \SI{0.35}{/V} under Sb-rich and Se-rich conditions, respectively (Fig. \ref{fig_eff}(a)).
%
%

The vital role of growth conditions in device performance has been widely reported by experiments, with selenisation treatment being proven effective in improving the conversion efficiency of \ce{Sb2Se3} by reducing the concentration of \textit{V}$_\mathrm{Se}$\cite{leng2014selenization,li2017sb2se3}. 
Thus, we study the \textit{V}$\mathrm{_{OC}}$ loss due to non-radiative recombination ($\Delta$\textit{V}$\mathrm{_{OC}^\mathrm{non-rad}}$) as a function of the growth condition (Fig. \ref{fig_eff}(b)). 
The largest $\Delta$\textit{V}$\mathrm{_{OC}^\mathrm{non-rad}}$ of \SI{0.41}{/V} is obtained under Sb-rich/Se-poor conditions. With the increase of the Se chemical potential $\mu_\mathrm{Se}$ (corresponding to more Se-rich conditions), $\Delta$\textit{V}$\mathrm{_{OC}^\mathrm{non-rad}}$ initially decreases until reaching a minimum (with the lowest $\Delta$\textit{V}$\mathrm{_{OC}^\mathrm{non-rad}}$ of \SI{0.14}{/V} achieved under intermediate growth conditions), and then increasing once again until the Se-rich limit. Nevertheless, $\Delta$\textit{V}$\mathrm{_{OC}^\mathrm{non-rad}}$ is much smaller under Se-rich conditions (\SI{0.22}{/V}) than under Sb-rich conditions (\SI{0.41}{/V}). These observations agree well with experimental findings that post-selenisation can improve the efficiency\cite{leng2014selenization,li2017sb2se3,rijal2021influence}, and that efficiency degradation occurs under extremely Sb/Se-rich conditions\cite{kim2021importance}.

To further analyse the most detrimental defect species, we divide the contributions to $\Delta$\textit{V}$\mathrm{_{OC}^\mathrm{non-rad}}$ based on each defect. Considering that the total $\Delta$\textit{V}$\mathrm{_{OC}^\mathrm{non-rad}}$ is not exactly a simple sum of individual defect contributions (as these depend on the total recombination rate), we normalise the coloured areas in Fig. \ref{fig_eff}(b) by:
\begin{equation}
\Delta V^\mathrm{non-rad}_\mathrm{OC\textunderscore contribution}=\frac{\Delta V^\mathrm{non-rad}_\mathrm{OC\textunderscore individual}}{\sum\Delta V^\mathrm{non-rad}_\mathrm{OC\textunderscore individual}}\times \Delta V^\mathrm{non-rad}_\mathrm{OC\textunderscore total}
\end{equation}

As shown in Fig. \ref{fig_eff}(b), we find that the conversion efficiency of \ce{Sb2Se3} is limited by vacancies, whereas antisites have a negligible impact on non-radiative recombination. 
This calls into question the prevailing assumption of antisites being the most detrimental defects to \ce{Sb2Se3} solar cell performance\cite{savory2019complex,duan2022sb2se3,zhao2022regulating}. Indeed, the concentrations of antisites are highest among all defect species (Fig. \ref{fig_conc}(c)), and they do introduce deep defect levels in the band gap (Fig. \ref{fig_tl}). Nevertheless, our calculated low to moderate non-radiative carrier capture coefficients (Fig. S4) suggest that antisites are benign with low recombination rates.

Among all vacancies, \textit{V}$_\mathrm{Se(2)}$ and \textit{V}$_\mathrm{Sb(1)}$ contribute most to $\Delta$\textit{V}$\mathrm{_{OC}^\mathrm{non-rad}}$ under Sb-rich and Se-rich conditions respectively (Fig. \ref{fig_eff}(b)), indicating that these defect species should be avoided to improve the photo-conversion efficiency in \ce{Sb2Se3}.
\textit{V}$\mathrm{_{Se(2)}}$ in particular is the most detrimental defect species due to its high defect concentration and large carrier capture coefficients for both electron and hole capture (Fig. \ref{fig_conc}(c) \& Table \ref{table_pes}), while \textit{V}$\mathrm{_{Se(1)}}$ and \textit{V}$\mathrm{_{Se(3)}}$ are found to have negligible impacts on efficiency.
This highlights the sensitivity of carrier trapping and recombination to small changes in structures/energetics and, consequently, the significant variation in behaviour that different inequivalent sites of the same nominal defect (e.g. selenium vacancies; \textit{V}$\mathrm{_{Se}}$) can exhibit.
The \ac{PESs} and calculated capture coefficients for the other two inequivalent sites of \textit{V}$\mathrm{_{Se}}$ are shown in Fig. S5 and Table S1, respectively. 

Using the aTLC model\cite{kim2020upper,kim2021ab}, the upper limit to conversion efficiency in \ce{Sb2Se3} solar cell is predicted as shown in Fig. \ref{fig_eff}(c).
Considering that the control of film orientation has been widely reported to improve the conversion efficiency of \ce{Sb2Se3} solar cells\cite{zhou2015thin,li2019orientation,lin2022crystallographic}, the directionally-dependent (anisotropic) conversion efficiency is also calculated based on the respective optical absorption coefficients (Fig. S18).
The orientation dependence of efficiency is calculated by considering unpolarised sunlight incident along each of the three crystallographic directions in \ce{Sb2Se3}.
We predict that the highest trap-limited conversion efficiency of \SI{25.1}{\percent} can be achieved along the [100] direction (which is the direction along the quasi-\ac{1D} [Sb$_4$X$_6$]$_n$ ribbons) under the optimal Se-moderate growth conditions. 
Experiments have also found \ce{Sb2Se3} films with controlled orientation along this direction to maximise device efficiencies\cite{zhou2015thin,li20199}.
Under the same conditions, the maximum difference in efficiency along different directions is \SI{1.5}{\percent}.
These results are calculated based on equilibrium defect concentrations at an annealing temperature of \SI{550}{\kelvin}, matching previous theoretical studies\cite{huang2021more} and representing an average value of the range used in the synthesis of champion \ce{Sb2Se3} devices\cite{zhao2022regulating,duan2022sb2se3}.
A high annealing temperature of \SI{648}{\kelvin} is reported to benefit the crystalline quality in the highest-efficiency \ce{Sb2Se3} solar cell\cite{zhao2022regulating}.
However, a higher annealing temperature will increase defect concentrations (assuming equilibrium under annealing) and thus further reduce the trap-limited conversion efficiency (Fig. S19).
We note that the effects of mobility and surface/interface recombination are not considered, which could also contribute to \textit{V}$\mathrm{_{OC}}$ loss in practical devices.


\subsection*{Extrinsic passivation of deep defects}
To investigate potential strategies for reducing the impact of selenium vacancies in \ce{Sb2Se3}, we further study the effect of oxygen substitution.
The focus on O$_\mathrm{Se}$ is inspired by the experimental observation that oxygen exposure is beneficial to \ce{Sb2Se3} solar cell performance,\cite{fleck2020oxygen,ren2023study} as well as our calculated result that the conversion efficiency of \ce{Sb2Se3} is largely limited by \textit{V}$_\mathrm{Se}$. 
As shown in Fig. \ref{fig_eff}(b), \textit{V}$_\mathrm{Se}$ is the only intrinsic point defect species found to significantly lower the efficiency under Sb-rich conditions, and also plays an important role even under Se-rich conditions due to its relatively high concentration (\textgreater \SI{e12}{\per\cubic\cm}). 
Thus, it is intuitive to surmise that detrimental Se vacancies could be passivated by \ce{O2} upon oxygen exposure.

\begin{figure*}[ht!]
    \centering   {\includegraphics[width=\textwidth]{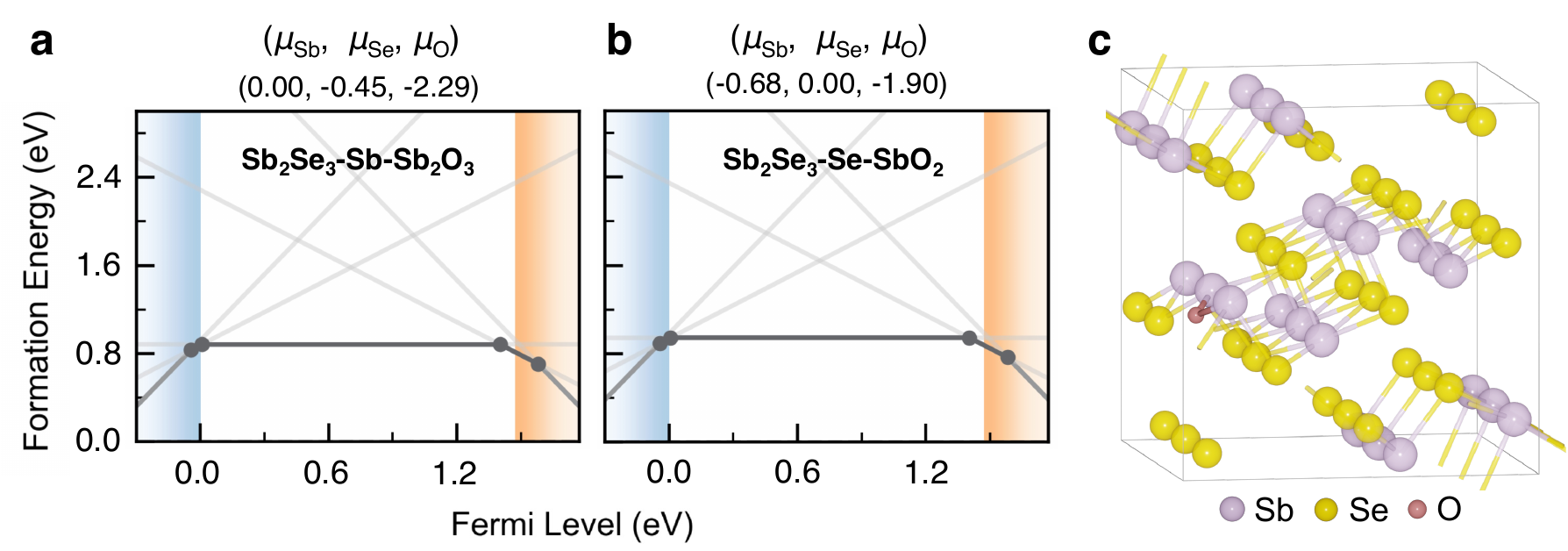}} \\
    \caption{(a)-(b) Formation energies of O$_\mathrm{Se(2)}$ in \ce{Sb2Se3} under (a) Sb-rich (with \ce{Sb2O3} being the oxygen-limiting phase) and (b) Se-rich (with \ce{SbO2} being the oxygen-limiting phase) conditions. The dark grey lines indicate energetically most favourable charge states. (c) Defect configuration of O$_\mathrm{Se(2)}^0$.}
    \label{fig_bulk_O}
\end{figure*}

To test our hypothesis and understand the role of oxygen, the structural configuration and energy of O$_\mathrm{Se}$ formation are studied. 
We mainly focus on the 2nd inequivalent site of Se as \textit{V}$_\mathrm{Se(2)}$ are responsible for $\textgreater$ \SI{99}{\percent} of the contribution to $\Delta$\textit{V}$_\mathrm{OC}$ among all Se vacancies.
As shown in Fig. \ref{fig_bulk_O}, the neutral state of O$_\mathrm{Se(2)}$ is thermodynamically stable across almost the entire band gap, leading to shallow defect levels --- which are inactive for recombination.
Moreover, the formation energies of O$_\mathrm{Se(2)}^0$ under O-poor conditions are relatively low (\SI{\sim 0.8}{\electronvolt}, which is similar to the formation energy of \textit{V}$_\mathrm{Se(2)}$). 
These results suggest the role of oxygen in passivating Se vacancies and eliminating their detrimental effects by shifting the deep recombination-active levels of \textit{V}$_\mathrm{Se(2)}$ (Fig. \ref{fig_tl}) to shallow inactive ones.

While structurally complex crystals such as \ce{Sb2Se3} can support the formation of many types of point defects, we have shown that only a subset will have a significant equilibrium population.
Vacancies and antisites are shown to be the dominant defects in as \ce{Sb2Se3} with high concentrations (\textgreater \SI{e14}{\per\cubic\cm}), while the concentrations of interstitials are relatively low.
Furthermore, by considering the processes of electron and hole capture, the most detrimental defects can be identified based on first-principles calculations.
This approach yields the ability to predict an upper limit for light-to-electricity conversion efficiency in a solar cell based on the bulk properties of the absorber material. 
For \ce{Sb2Se3}, \textit{V}$_\mathrm{Se}$ and \textit{V}$_\mathrm{Sb}$ contribute most to \textit{V}$_\mathrm{OC}$ deficit under Sb-rich and Se-rich conditions, respectively.
The largest \textit{V}$_\mathrm{OC}$ deficit is predicted to be \SI{0.54}{\volt} under Sb-rich conditions with a conversion efficiency of 18\%.
Intermediate growth conditions result in lower concentrations of defects and thus higher conversion efficiency (25\%).
Therefore, to improve the device performance, it is critical to suppress vacancy formation and optimise the growth conditions.
As a proof of concept, we showed that the detrimental effects of Se vacancies can be reduced by oxygen passivation (i.e. the formation of O$_\mathrm{Se}$).
An alternative passivation strategy is Fermi level engineering, where an extrinsic donor dopant is introduced during crystal growth/annealing to push the Fermi level higher in the gap, increasing the formation energy of \textit{V}$\mathrm{_{Se}}^{2+}$ and thus reducing its concentration.
In conclusion, our work provides a microscopic understanding of the efficiency limit of \ce{Sb2Se3} solar cells.

\section*{Computational Procedures}

\subsection*{Trap-limited conversion efficiency}
The light-to-electricity conversion efficiency of a solar cell depends on the proportion of electron-hole pairs extracted from the absorber layer and is limited by different electron-hole recombination mechanisms. In this work, trap-limited conversion efficiency of a single junction solar cell is determined by considering radiative and non-radiative recombination processes following the methodology proposed by Kim et al.\cite{kim2020upper,kim2021ab}. The effects of band gap, (film) thickness-dependent optical absorption and defect properties are taken into account. The mobility of electron-hole pairs is assumed to be infinitely high, so scattering mechanisms are neglected. This assumption can be justified by the relatively high carrier mobility reported in \ce{Sb2Se3}\cite{wang2022band}. Recombination at surfaces and interfaces are beyond the scope of this work and thus not considered in the following.
~\\
~\\
\textbf{Radiative recombination.}
An excited electron in the \ac{CB} can recombine with a hole in the \ac{VB} and energy is released as photons. This is an unavoidable process known as radiative recombination which is an inverse process of light absorption. 
~\\
~\\
Under the assumption of ideal reflection at the bottom of the absorber, the photon absorptivity \textit{a} is calculated by
\begin{equation}
a(E;W) = 1 - e^{-2\alpha(E)W}
\end{equation}
where \textit{E} and \textit{W} are the photon energy and film thickness, respectively, and $\alpha$ is the optical absorption coefficient.
~\\
~\\
Assuming that each absorbed photon generates one electron-hole pair, the short-circuit current $J_\textrm{SC}$ is given by:
\begin{equation}
J_\textrm{SC}(W) = e\int_{E_g}^{\infty}a(E;W)\Phi_\textrm{sun}(E)\textrm{d}E
\end{equation}
where \textit{e} is the elementary charge, $\Phi_\textrm{sun}(E)$ is incident spectral photon flux density at the photon energy \textit{E}. Here a standard AM1.5 solar spectrum is considered. 
~\\
~\\
The radiative recombination rate $\textit{R}_\textrm{rad}$ at temperature \textit{T} and voltage \textit{V} is given by:
\begin{equation}
\begin{aligned}
R_\textrm{rad}(V) &= \frac{2\pi}{c^2h^3}\int_{0}^{\infty}a(E;W)[e^{\frac{E-eV}{k_BT}} - 1]^{-1} E^2\textrm{d}E 
\\
&\approx \frac{2\pi}{c^2h^3} e^{\frac{eV}{k_BT}}\int_{0}^{\infty}a(E;W)[e^{\frac{E}{k_BT}} - 1]^{-1} E^2\textrm{d}E
\end{aligned}
\end{equation}
~\\
The net current density $J^\textrm{rad}$ generated under illumination in the radiative limit is given by:
\begin{equation}
J^\textrm{rad}(V;W) = J_\textrm{SC}(W) + J_0^\textrm{rad}(W)(1-e^{\frac{eV}{k_\textrm{B}T}})
\end{equation}
where the saturation current $J_0^\textrm{rad}$ = \textit{e}$R_\textrm{rad}(0)$.
~\\
~\\
\textbf{Non-radiative recombination.}
The main cause of efficiency loss in a solar cell usually involves non-radiative recombination facilitated by deep-level defects. Identifying the detrimental defect species is thus crucial to improving the device performance.
~\\
~\\
\textit{Defect formation energy.}
The formation energy of a point defect \textit{D} in charge state \textit{q} is calculated by the equation\cite{zhang1991chemical,freysoldt2014first}:
\begin{equation}
    \Delta E^{\textit{f}}_{D,q} = E_{D,q} - E_\textrm{host} - \sum_{i}n_{i}\mu_{i} + qE_F +E_\textrm{corr}
\end{equation}
where $E_{D,q}$ and $E_\textrm{host}$ are the total energies of the supercells with and without the defect \textit{D}, respectively. $n_i$ and $\mu_{i}$ represent the number and the chemical potential of added ($n_i$ \textgreater { }0) or removed ($n_i$ \textless { }0) atom of type \textit{i}, respectively. $E_F$ is the Fermi level. $E_\textrm{corr}$ accounts for the finite-size corrections for charged defects under periodic boundary conditions. In this work, the correction scheme developed by Kumagai and Oba\cite{kumagai2014electrostatics} which accounts for anisotropic dielectric screening is employed, which has been extensively shown to be both accurate and robust\cite{walsh2021correcting,oba2018design}.

~\\
\textit{Defect transition level.}
The thermodynamic charge \ac{TL} $\varepsilon$(\textit{q}$_1$/\textit{q}$_2$) is defined as the Fermi-level position at which the formation energies of charge states \textit{q}$_1$ and \textit{q}$_2$ of a defect are the same, and can be obtained from the relation:
\begin{equation}
\varepsilon(\textit{q}_1/\textit{q}_2)=\frac{\Delta E^f_{D,q_1}(\textit{E}_\textit{F}=0)-\Delta E^f_{D,q_2}(\textit{E}_\textit{F}=0)}{\textit{q}_2-\textit{q}_1}
\end{equation}
where $\Delta E^{\textit{f}}_{D,q}$($\textit{E}_\textit{F}=0$) is the formation energy of a defect \textit{D} in the charge state \textit{q} when the Fermi level is at the \ac{VBM}.
~\\
~\\
\textit{Defect and charge carrier concentration.}
The self-consistent Fermi level is calculated by an iterative method\cite{squires2023py,buckeridge2019equilibrium} based on the charge neutrality condition
\begin{equation}
\sum_{D,q}qC_{D,q}-n_0+p_0=0
\end{equation}
where $C_{D,q}$ is the defect concentration of a defect \textit{D} in its accessible charge state \textit{q}. \textit{n}$_0$ and \textit{p}$_0$ are concentrations of free electrons and holes, respectively.
~\\
~\\
The defect concentration $C_{D,q}$ is given as
\begin{equation}
C_{D,q}=gN_{D}e^{\frac{- \Delta E^{\textit{f}}_{D,q}}{k_\textrm{B}T_\textrm{anneal}}}
\end{equation}
where \textit{g} is the degeneracy term including spin and geometry degeneracy. $N_{D}$ is the number of possible sites for defect \textit{D} to form in the supercell per volume. $k_\textrm{B}$ is the Boltzmann constant, and \textit{T}$_\textrm{anneal}$ is the temperature at which the host material is annealed/synthesised.
~\\
~\\
The electron (\textit{n}$_0$) and hole (\textit{p}$_0$) concentrations are obtained by
\begin{equation}
n_0=\int_{E_\textrm{CBM}}^{\infty}\rho(E)f(E)dE
\end{equation}
\begin{equation}
p_0=\int_{-\infty}^{E_\textrm{VBM}}\rho(E)[1-f(E)]dE
\end{equation}
where $\rho(E)$ is the \ac{DOS} per unit volume, and $f(E)$ is the Fermi-Dirac distribution function which represents the likelihood of an electron occupying an energy state \textit{E}
\begin{equation}
f(E) = \frac{1}{e^{1+\frac{E-E_F}{k_{B}T}}}
\end{equation}
where $E_F$ is the Fermi level, and \textit{T} is the temperature.
~\\
~\\
It is worth noting that \textit{T}$_\textrm{anneal}$ and \textit{T} are specified as different values in this work. This is due to the fact that after rapid quenching, defects usually become `frozen in' at annealing temperature by kinetic barriers. Therefore, the defect concentrations are fixed at \textit{T}$_\textrm{anneal}$, while the change of charge states of the same defect species is possible, and the concentrations of free electrons and holes are able to re-equilibrate until the charge neutrality condition is met at the measurement temperature \textit{T}.
~\\
~\\
\textit{Carrier capture coefficient.}
The non-radiative carrier capture for deep-level defects can be accurately simulated via \ac{MPE} from first-principles calculations\cite{alkauskas2014first}. Within the framework of \ac{MPE}, the transition between delocalised bulk state and localised defect state can be treated by considering first order of electron-phonon coupling perturbation. 
The carrier capture coefficient is determined by Fermi's golden rule
\begin{equation}
C=\frac{2\pi}{\hbar}Vg\sum_{m}\omega_{m}\sum_{n}|\Delta H^{\textrm{e-ph}}_{im;fn}|^2\delta(E_{im}-E_{fn})
\end{equation}
where \textit{V} is the volume of the supercell. \textit{g} is the degeneracy term accounting for the number of equivalent transition pathways, which includes spin and geometry degeneracy\cite{kavanagh2022impact}. \textit{n} and \textit{m} are quantum numbers of ionic states. $\omega_{m}$ represents the thermal occupation. $E_{\{im,fn\}}$ are total energies. $\Delta H^{\textrm{e-ph}}_{im;fn}$ is the electron-phonon coupling matrix element. Under the linear-coupling approximation, the matrix element is determined by Taylor expansion in {\textit{Q}} around {\textit{Q}$_0$} with only the first-order terms preserved
\begin{equation}
\Delta H^{\textrm{e-ph}}_{im;fn}=\sum_{k}\braket{\Psi_{i}|\partial \hat{H}/\partial Q_{k}|\Psi_{f}}\braket{\chi_{im}|Q_{k}-Q_{0;k}|\chi_{fn}}
\end{equation}
where $\Psi_{\{i,f\}}$ are the many-body electronic wavefunctions, and $\chi_{\{im,fn\}}$ are the ionic wavefunctions. It sums all phonon modes $Q_k$.
~\\
~\\
Under the effective-\ac{1D} approximation, one effective phonon mode which represents the strongest interaction with the deformation of defect configuration is used. Using Kohn-Sham \ac{DFT}, many-body Hamiltonian and wavefunctions are replaced by single-particle ones. Thus, the carrier capture coefficient is calculated by
\begin{equation}
\tilde{C}=\frac{2\pi}{\hbar}Vg\sum_{m}\omega_{m}\braket{\psi_{i}|\partial \hat{h}/\partial Q|\psi_{f}}\sum_{n}|\braket{\chi_{im}|Q-Q_{0}|\chi_{fn}}|^2\delta(\Delta E+m\hbar \Omega_i -n\hbar \Omega_f)
\end{equation}
where $\hbar$ and $\psi_{\{i,f\}}$ are the single-particle Hamiltonian and single-particle wavefunctions, respectively. $\Omega_{\{i,f\}}$ are the phonon frequencies of initial and final states. 
A one dimensional generalized coordinate \textit{Q} is used to represent atomic deformation \cite{stoneham1975theory}, which is defined as 
\begin{equation}
Q^2=\sum_{\alpha}M_{\alpha}\Delta R^2_{\alpha}
\end{equation}
where $M_{\alpha}$ and $\Delta R_{\alpha}$ are the mass and the displacement between the initial and final states of an atom $\alpha$, respectively.
~\\
~\\
Two types of scaling parameters are considered in this work to correct the capture coefficients when necessary and were calculated using \textsc{Nonrad} \cite{turiansky_nonrad_2021}. In the cases of carrier captured by a charged defect, Sommerfeld parameter \textit{s}(\textit{T}) \cite{passler1976relationships} is calculated to account for the Coulombic interaction between the delocalised carrier and charged defect. While when a charged defect supercell is used to calculate the electron-phonon matrix elements, a scaling factor \textit{f} is calculated to correct the charge density near the defect. 
After taking into account the scaling parameters, the carrier capture coefficient is given as:
\begin{equation}
C=s(T)f\tilde{C}
\end{equation}
~\\
The capture cross-section $\sigma$ is given by
\begin{equation}
\sigma=\frac{C}{\langle v \rangle}
\end{equation}
where $\langle v \rangle$ = $\sqrt{3k_BT/m^*}$ is the average thermal velocity of the carrier. $m^*$ is the average effective mass and in this work, $m_e^*$=0.35 and $m_h^*$=0.90\cite{wang2022lone}.
~\\
~\\
\textit{Non-radiative recombination rate.}
For non-degenerate semiconductors, non-radiative recombination rate $R_\textrm{SRH}$ under steady-state conditions is calculated based on \ac{SRH} statistics\cite{shockley1952statistics,hall1952electron}
\begin{equation}
R_\textrm{SRH}=\frac{np-n_{0}p_{0}}{(n+n_{1})\tau_{p}+(p+p_{1})\tau_{n}}
\end{equation}
where \textit{n} and \textit{p} are concentrations of electrons and holes, respectively. \textit{n}$_0$ and \textit{p}$_0$ are concentrations of electrons and holes at thermal equilibrium, respectively. $n_1$ = $N_c$exp($\frac{E_t-E_\textrm{CBM}}{k_BT}$) and $p_1$ = $N_v$exp($\frac{E_\textrm{VBM}-E_t}{k_BT}$) are electron and hole densities when the Fermi level lies at the trap level $E_t$, and $N_c$ and $N_v$ are effective density of states for \ac{CB} and \ac{VB}, respectively. $\tau_{p}$ and $\tau_{n}$ are lifetime for hole and electron capture, respectively, which are given by
\begin{equation}
\begin{aligned}
\tau_{p}&=\frac{1}{N_TC_p} \\
\tau_{n}&=\frac{1}{N_TC_n}
\end{aligned}
\end{equation}
where $C_p$ and $C_n$ are hole and electron capture coefficients, respectively. $N_T$ is the total defect concentration. 
The relative defect concentration for a defect \textit{D} in a certain charge state \textit{q} is calculated under kinetic equilibrium. For example, in the transitions between $q$, $q-1$ and $q-2$ charge states: 
 \begin{equation}
 \ce{
 \textit{D}\,^q <=>[e^-][h^+]
 \textit{D}\,^{q-1} <=>[e^-][h^+] \textit{D}\,^{q-2}}
 \end{equation}
Under steady-state conditions (constant illumination), the net electron capture rate by $\textit{D}\,^{q}$($\textit{D}\,^{q-1}$) should be equal to the net hole capture rate by $\textit{D}\,^{q-1}$($\textit{D}\,^{q-2}$).
Considering in \ce{Sb2Se3}, the equilibrium carrier density $n_0$ and $p_0$ is much lower than the photo-generated carrier density $\Delta n$, and carrier emission is assumed to be negligible for deep-level defects\cite{alkauskas2016role}:
\begin{equation}
\begin{aligned}
N^{q}C^{q}_n \Delta n = N^{q-1}C^{q-1}_p \Delta n  \\
N^{q-1}C^{q-1}_n \Delta n = N^{q-2}C^{q-2}_p \Delta n
\end{aligned}
\end{equation}
The sum of concentrations of $D^q$, $D^{q-1}$ and $D^{q-2}$ is kept fixed and determined by the concentration of \textit{D} at thermodynamic equilibrium ($N_\mathrm{tot}$).
The relative defect concentrations are then calculated as:
\begin{equation}
\begin{aligned}
N^{q-1}=\frac{N_\mathrm{tot}}{1+\frac{C^{q-1}_p}{C^{q}_n}+\frac{C^{q-1}_n}{C^{q-2}_p}}     \\
N^{q}=N^{q-1}\frac{C^{q-1}_p}{C^{q}_n}     \\
N^{q-2}=N^{q-1}\frac{C^{q-1}_n}{C^{q-2}_p}
\end{aligned}
\end{equation}
The total SRH recombination rate $R_\textrm{SRH}$ is the sum of recombination rates for all defect charge states.
~\\ 
~\\
Upon illumination, there is an extra contribution of photogenerated carrier concentration $\Delta n$ which is given by
\begin{equation}
\Delta n=\frac{1}{2}[-n_0-p_0+\sqrt{(n_0+p_0)^2-4n_0p_0(1-e^{\frac{eV}{k_BT})}}]
\end{equation}
~\\
Consequently, the concentrations of electrons \textit{n} and holes \textit{p} are calculated by
\begin{equation}
\begin{aligned}
n&=n_0+\Delta n \\
p&=p_0+\Delta n
\end{aligned}
\end{equation}
~\\
\textit{Trap-limited conversion efficiency.}
By including both radiative and non-radiative recombination, the net current density \textit{J} under a bias voltage \textit{V} is defined as
\begin{equation}
J(V;W) = J_\textrm{SC}(W) + J_0^\textrm{rad}(W)(1-e^{\frac{eV}{k_\textrm{B}T}})-eR_\textrm{SRH}(V)W
\end{equation}
~\\
The maximum efficiency is defined as the ratio of the maximum power density to the incident light power density, which is given by:
\begin{equation}
\eta_{max} = \textrm{max}_V(\frac{JV}{e\int_{0}^{\infty}E\Phi_\textrm{sun}(E)\textrm{d}E})
\end{equation}
~\\
\subsection*{First-principles calculations}

All calculations for the underling total energies were performed based on Kohn-Sham \ac{DFT}\cite{kohn1965self,dreizler1990density} as implemented in \ac{VASP}\cite{kresse1996efficient}. 
The \ac{PAW} method\cite{kresse1999ultrasoft} was employed with converged plane-wave energy cutoffs of \SI{350}{\electronvolt} and \SI{400}{\electronvolt} for \ce{Sb2Se3} with intrinsic defects and extrinsic oxygen defects, respectively.
Both structural relaxation and static calculations of the pristine structure and defects in \ce{Sb2Se3} were performed using the Heyd-Scuseria-Ernzerhof hybrid exchange-correlation functional (HSE06)\cite{heyd2003hybrid,krukau2006influence} and the D3 dispersion correction\cite{grimme2004accurate}, which have been demonstrated to well reproduce the geometric and electronic properties in \ce{Sb2Se3}\cite{wang2022lone}.
We note that electron-phonon coupling has been shown to impact the band gap of antimony chalcogenides,\cite{liu2023strong} however the full inclusion of these effects is beyond current computational capabilities and they are not expected to dramatically alter the conclusions here.
~\\
~\\
\textbf{Bulk crystal modelling.}
The bulk structure calculation of \ce{Sb2Se3} was carried out using a unit cell containing twenty atoms with 15$\times$6$\times$6 $\varGamma$-centred Monkhorst-Pack \textit{k}-point mesh. The atomic positions were optimised until the Hellman-Feynman forces on each atom were below \SI{5e-4}{eV/\angstrom}.
~\\
~\\
\textbf{Defect modelling.}
The \textsc{doped} Python package\cite{kavanagh_2023_8403524} was used for the generation, parsing and analysis/plotting of defect supercell calculations. 
For all calculations of defects, the convergence criterion of forces on each atom was set to \SI{0.01}{eV/\angstrom}. Spin polarisation was turned on during the geometry relaxation. 
A 3$\times$1$\times$1 (\SI{11.86}{\angstrom}$\times$\SI{11.55}{\angstrom}$\times$\SI{11.93}{\angstrom}) 60-atom supercell and 2$\times$2$\times$2 $\varGamma$-centred \textit{k}-point mesh was used for both geometry optimisation and static calculations. 
The \textsc{ShakeNBreak}\cite{mosquera2022shakenbreak} global structure searching method was used to aid the identification of ground-state defect geometries.
Initial defect configurations were obtained by local bond distortions (of both compression and stretching between 0\% and 60\% with 10\% as an interval) around the defect and random displacement (\textit{d}) to all atoms in the supercell, which has been proved to efficiently map complex defect \ac{PESs} and identify ground state structures\cite{mosquera2023identifying}. 
\textit{d} is stochastically selected from a normal distribution of a standard deviation $\sigma$,
\begin{equation}
d \leftarrow \frac{1}{\sigma\sqrt{2\pi}}\textrm{exp}(-\frac{d^2}{2\sigma^2})
\end{equation}
$\sigma$ = 0.25, 0.20 and 0.15 Å were tested.
~\\
~\\
The defect and charge carrier concentrations under thermodynamic equilibrium are calculated using \textsc{py-sc-fermi}\cite{squires2023py,buckeridge2019equilibrium}.
Anharmonic carrier capture coefficients were calculated using \textsc{CarrierCapture.ji}\cite{kim2020carriercapture}.
~\\
\bmhead{Supplemental information}
Supplemental information can be found online.
~\\
\bmhead{Acknowledgments}
X.W. thanks Sunghyun Kim, Menglin Huang, and Shiyou Chen for valuable discussions on carrier capture processes. Via our membership of the UK's HEC Materials Chemistry Consortium, which is funded by EPSRC (EP/X035859/1), this work used the ARCHER2 UK National Supercomputing Service (http://www.archer2.ac.uk). 
This work was supported by EPSRC project EP/X037754/1.
X.W. acknowledges Imperial College London for support via a President's PhD Scholarship. 
S.R.K. acknowledges funding from Samsung Electronics Ltd. (SAIT).
~\\
\bmhead{Declaration of interests}
The authors declare no competing interests.


\bibliography{references}

\end{document}


\begin{acronym}
\acro{CC}{configuration coordinate}
\acro{PESs}{potential energy surfaces}
\acro{PES}{potential energy surface}
\acro{TLs}{transition levels}
\acro{TL}{transition level}
\acro{CBM}{conduction band minimum}
\end{acronym}

\newpage

\section{S1. Defect configurations of interstitials}

\begin{figure}[h!]
    \centering    {\includegraphics[width=0.55\textwidth]{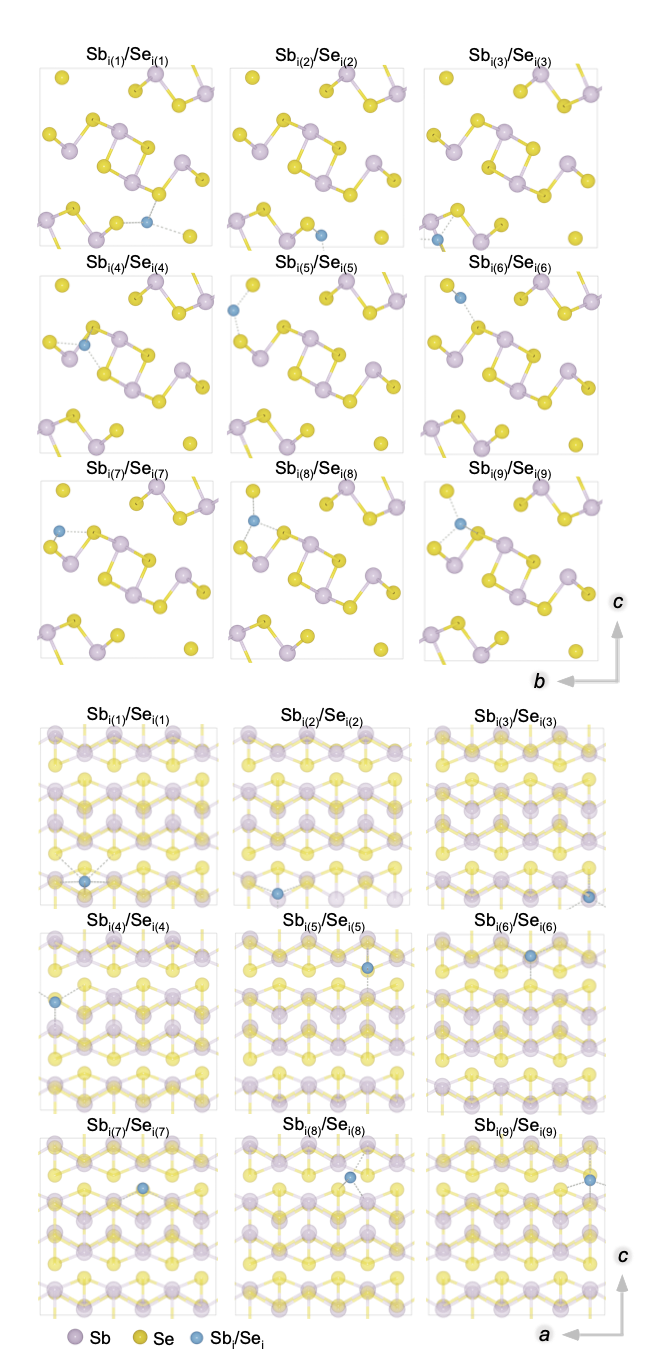}} \\
    \caption{Nine initial configurations of Sb/Se interstitials considered in this work from different perspectives.}
    \label{fig_int}
\end{figure}

\newpage

\section{S2. Formation energies under Se-moderate conditions}
\begin{figure}[h]
    \centering    {\includegraphics[width=0.8\textwidth]{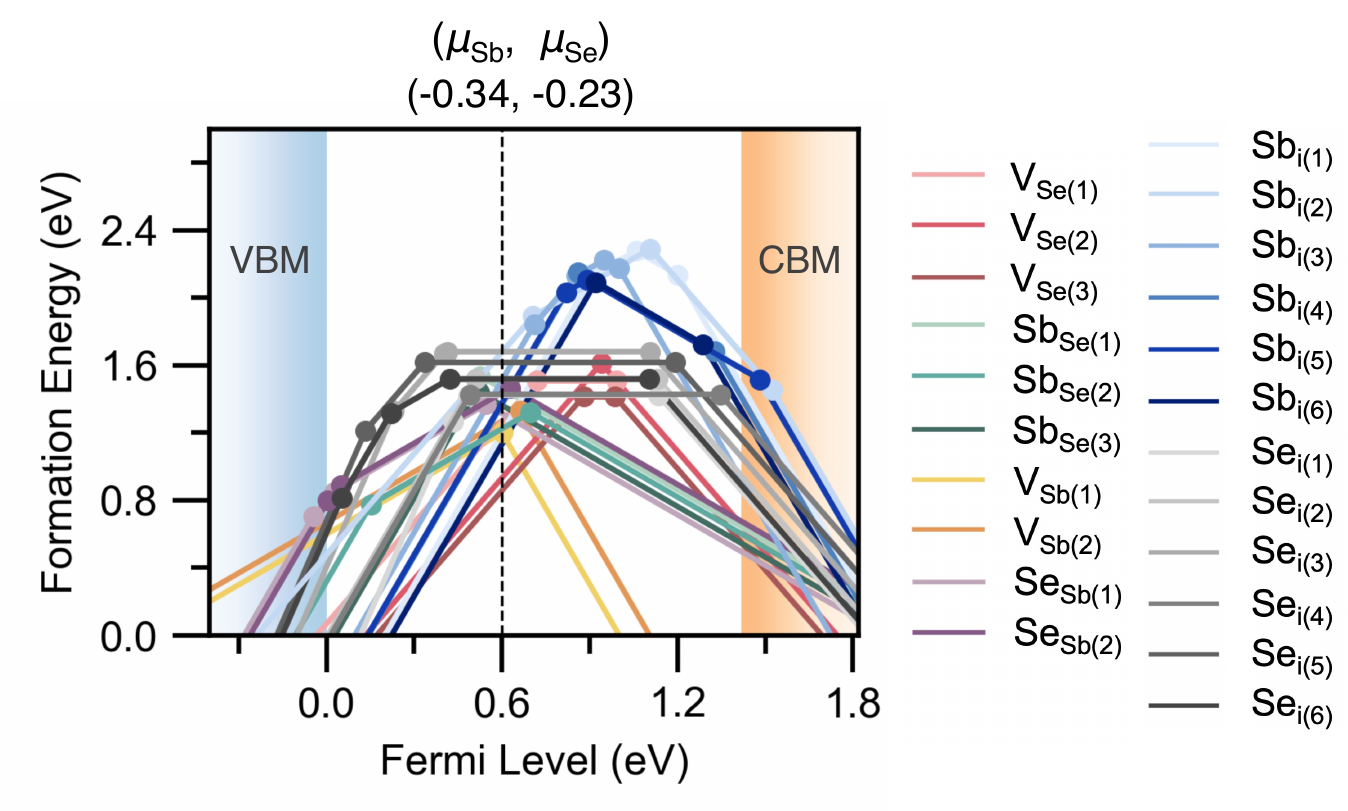}} \\
    \caption{Formation energies of all intrinsic point defects in \ce{Sb2Se3} under Se-moderate conditions. The dashed line indicates self-consistent Fermi level at \SI{300}{\kelvin} in \ce{Sb2Se3} crystals grown at \SI{550}{\kelvin}.}
    \label{fig_carrier_conc}
\end{figure}
\section{S3. Carrier concentration}
\begin{figure}[h]
    \centering    {\includegraphics[width=0.4\textwidth]{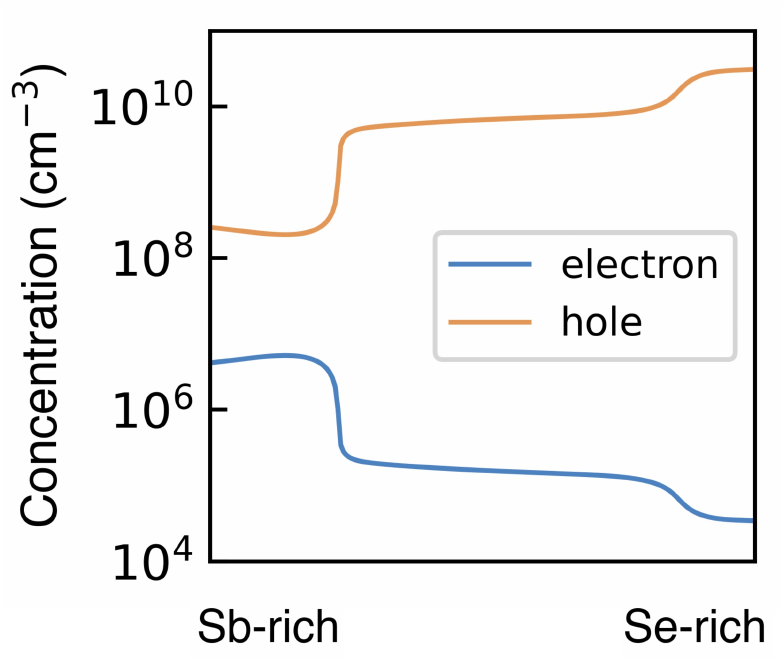}} \\
    \caption{Carrier concentration in the dark at \SI{300}{\kelvin} in \ce{Sb2Se3} crystals grown at \SI{550}{\kelvin} as a function of the growth condition.}
    \label{fig_carrier_conc}
\end{figure}

\newpage

\section{S4. Carrier capture processes}
\begin{figure}[h!]
    \centering    {\includegraphics[width=\textwidth]{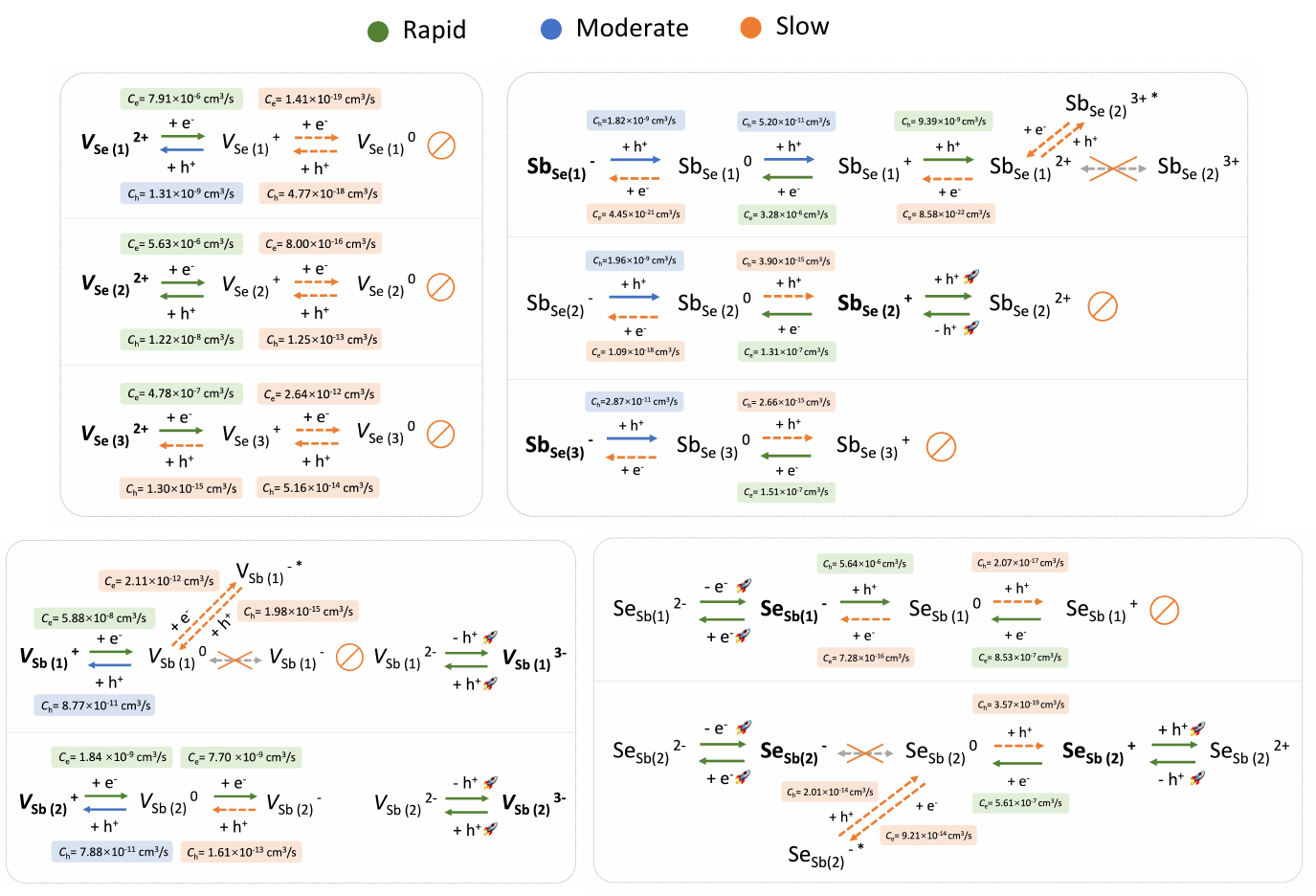}} \\
    \caption{Pathways for trap-mediated electron and hole capture. Defect species in bold are thermodynamically stable states at calculated self-consistent Fermi levels, which are the most likely starting points in capture processes. The defect species with superscript asterisks refer to metastable defect configurations. $C_e$ and $C_h$ are electron and hole capture coefficients, respectively. Green, blue and orange colours indicate rapid, intermediate and slow capture. Capture coefficients smaller than \SI{e-25}{{\cubic\cm\per\second}} are not shown. `$\oslash$' refers to transitions from states with extremely low predicted concentrations under illumination (see text for details). Transitions with large mass-weighted displacements are also ruled out, as indicated by an `X' mark.}
    \label{fig_path}
\end{figure}
The electron and hole capture processes for defects with high concentrations (i.e. vacancies and antisites) are investigated. 
For each defect \textit{D}, we start from the thermodynamically stable state \textit{q} at the calculated self-consistent Fermi level, and consider the single-electron transition between $D^\textit{q}$ and $D^{\textit{q}-1}$/$D^{\textit{q}+1}$. 
If the transition is fast enough and the concentration of $D^{\textit{q}-1}$/$D^{\textit{q}+1}$ is relatively high, we further consider the neighbouring transitions of $D^{\textit{q}-1}$ $\leftrightarrow$ $D^{\textit{q}-2}$ / $D^{\textit{q}+1}$ $\leftrightarrow$ $D^{\textit{q}+2}$. The relative defect concentrations are obtained under steady-state conditions (shown in Method).
Within this approach, if any state has an extremely low concentration (smaller than \SI{e-25}{{\cubic\cm\per\second}}), the remaining neighbouring transitions are not considered (labelled as `$\oslash$' in Fig. \ref{fig_path})

The `X' mark indicates the transition is ruled out due to significant dissimilarity between two states with a large mass-weighted displacement ($\Delta$\textit{Q}). In such cases, metastable defect configurations should be considered as potential intermediate configurations (example shown in S5.2).\cite{kavanagh2022impact,alkauskas2016role}

The rocket symbol indicates the defect level is shallow. If the defect level is close to or inside the band edge, it will be very rapid for the defect to both capture and emit a carrier, and thus unlikely for such defect to be an effective recombination centre. Here double thermal energy at room temperature (i.e. 50 meV) was used as a criterion for considering a defect level to be shallow --- if the energy difference between the defect level and band edge is smaller than 50 meV, we excluded that transition. ~\\

\subsection{S5.1 Vacancies}

\subsubsection{S5.1.1 \textit{V}$^{2+}_\mathrm{Se}$ $\leftrightarrow$ \textit{V}$^{+}_\mathrm{Se}$}

\begin{figure}[h!]
    \centering    {\includegraphics[width=0.92\textwidth]{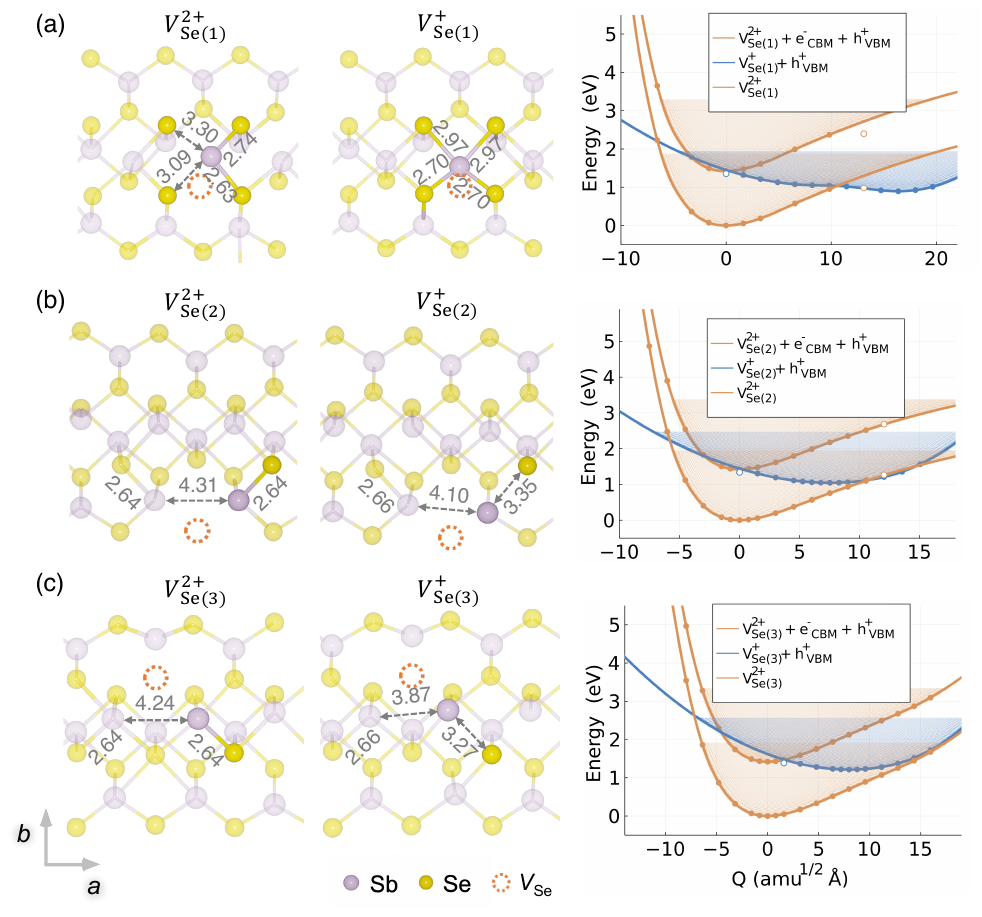}} \\
    \caption{Defect configurations and one-dimensional configuration coordinate diagrams for the charge transition between \textit{V}$^{2+}_\mathrm{Se}$ and \textit{V}$^{+}_\mathrm{Se}$ in \ce{Sb2Se3} for inequivalent (a) site 1, (b) site 2 and (c) site 3. Solid circles are datapoints obtained by DFT calculations and used for fitting, while hollow circles are discarded for fitting due to energy-level crossings. Solid lines represent the best fits to the data.}
    \label{fig_vse}
\end{figure}

\begin{table*}[ht!]
\caption{Key parameters used to calculate the carrier capture coefficients in the transition of \textit{V}$^{2+}_\mathrm{Se}$ $\leftrightarrow$ \textit{V}$^{+}_\mathrm{Se}$ for different inequivalent sites: mass-weighted distortion $\Delta$\textit{Q} (amu$^{1/2}$\AA), energy barrier $\Delta$\textit{E}$_\textrm{b}$  (meV), degeneracy factor \textit{g} of the final state, electron-phonon coupling matrix element $W_{if}$ and scaling factor \textit{s}(\textit{T})\textit{f} at \SI{300}{\kelvin}; and calculated capture coefficient \textit{C} (\SI{}{\cubic\cm\per\second}) and capture cross section $\sigma$  (\SI{}{\cm\squared}) at \SI{300}{\kelvin}}
\label{table_vse}
\begin{tabular*}{\textwidth}{@{\extracolsep{\fill}}c@{\extracolsep{\fill}}c@{\extracolsep{\fill}}c@{\extracolsep{\fill}}c@{\extracolsep{\fill}}c@{\extracolsep{\fill}}c@{\extracolsep{\fill}}c@{\extracolsep{\fill}}c@{\extracolsep{\fill}}c}
    \hline
Species & $\Delta$\textit{Q} & \begin{tabular}[c]{@{}c@{}}Capture\\ process\end{tabular} & $\Delta$\textit{E}$_\textrm{b}$ & \textit{g} & $W_{if}$ & \textit{s}(\textit{T})\textit{f} & \textit{C} & $\sigma$ \\     \hline
\multirow{2}{*}{\textit{V}$_\mathrm{Se(1)}$} &\multirow{2}{*}{16.37} & Electron & \SI{2}{} & 2 & \SI{3.74e-2}{} & 1.73  &\SI{7.91e-6}{} & \SI{4.01e-13}{} \\
 & & Hole & \SI{138}{} & 1 & \SI{2.28e-2}{} & 0.38 &\SI{1.31e-9}{} & \SI{1.07e-16}{} \\     \hline
\multirow{2}{*}{\textit{V}$_\mathrm{Se(2)}$} &\multirow{2}{*}{7.52} & Electron & \SI{2}{} & 4 & \SI{1.81e-2}{} & 2.09  &\SI{5.63e-6}{} & \SI{2.85e-13}{} \\
 & & Hole & \SI{83}{} & 1 & \SI{1.76e-2}{} & 0.35 &\SI{1.22e-8}{} & \SI{9.89e-16}{} \\     \hline
 \multirow{2}{*}{\textit{V}$_\mathrm{Se(3)}$} &\multirow{2}{*}{7.98} & Electron & \SI{53}{} & 4 & \SI{1.44e-2}{} &  1.80 &\SI{4.78e-7}{} & \SI{1.30e-15}{} \\
 & & Hole & \SI{1423}{} & 1 & \SI{6.86e-3}{} & 0.53 &\SI{1.30e-15}{} & \SI{1.05e-22}{} \\     \hline
\end{tabular*}
\end{table*}

\begin{figure}[h!]
    \centering    {\includegraphics[width=0.78\textwidth]{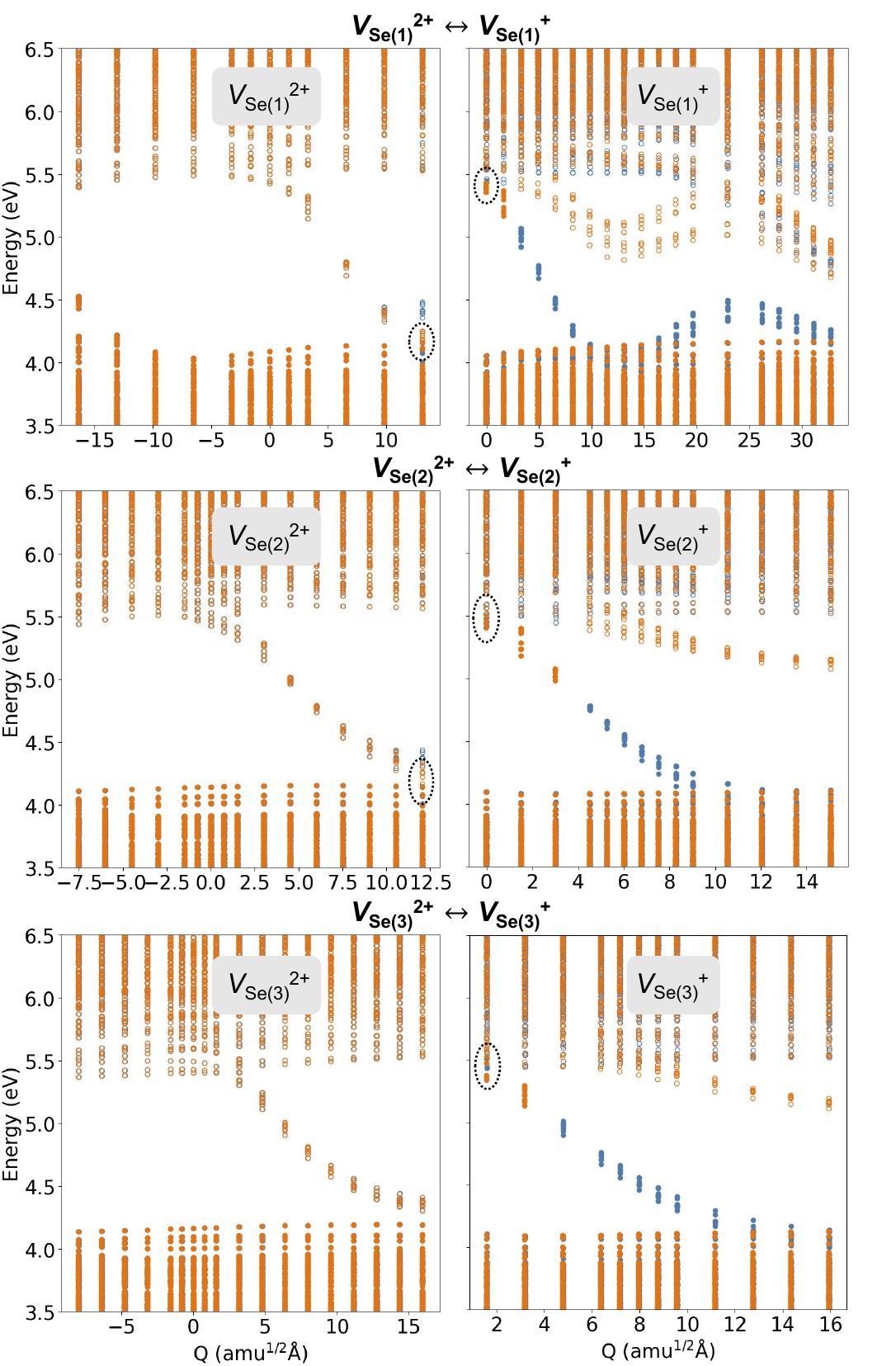}} \\
    \caption{Evolution of Kohn-Sham eigenstates of \textit{V}$\mathrm{_{Se}}^{2+}$ and \textit{V}$\mathrm{_{Se}}^{+}$ as a function of the structural deformation \textit{Q} for all inequivalent sites. Solid and hollow circles refer to occupied and unoccupied states, respectively. Different colours represent different spin channels. Dotted ovals represent crossing points which are discarded for fitting.}
    \label{fig_vse_eig}
\end{figure}

As shown in Fig. \ref{fig_path}, for Se vacancies, the transitions between \textit{V}$^{2+}_\mathrm{Se}$ and \textit{V}$^{+}_\mathrm{Se}$ have larger capture coefficients than the other transitions. 
Thus, here we show the defect configurations and \ac{PESs} of \textit{V}$^{2+}_\mathrm{Se}$ $\leftrightarrow$ \textit{V}$^{+}_\mathrm{Se}$ for all inequivalent sites (Fig. \ref{fig_vse}). 
All electronic eigenstates as a function of \textit{Q} were carefully checked and datapoints where the occupation of single-particle defect levels changed due to crossing the band edges are not used for fitting (Fig. \ref{fig_vse_eig}).
For all these three sites, the main structural deformation during transition is related to the relaxation of one neighbouring Sb atom around \textit{V}$_\mathrm{Se}$. This results in the elongation/compression of two Sb-Se bonds in \textit{V}$_\mathrm{Se(1)}$, and one Sb-Se bond in \textit{V}$_\mathrm{Se(2)}$ and \textit{V}$_\mathrm{Se(3)}$ (highlighted in Fig. \ref{fig_vse}). Consequently, a larger $\Delta$\textit{Q} of 16.37 amu$^{1/2}$\AA{ }is obtained for \textit{V}$^{2+}_\mathrm{Se(1)}$ $\leftrightarrow$ \textit{V}$^{+}_\mathrm{Se(1)}$ (Table \ref{table_vse}).
Although the shapes of \ac{PESs} and electron-phonon coupling $W_{if}$ are similar for these three sites, the calculated capture coefficients can be quite different as they are sensitive to the positions of \ac{TLs}.
Similar (2+/+) \ac{TLs} (of 0.9 and 1.0 eV for \textit{V}$_\mathrm{Se(1)}$ and \textit{V}$_\mathrm{Se(2)}$, respectively) which are close to the middle of the band gap lead to similar large capture coefficients for \textit{V}$_\mathrm{Se(1)}$ and \textit{V}$_\mathrm{Se(2)}$ (on the order of \SI{e-6}{} and \SIrange{e-8}{e-9}{\cubic\cm\per\second} for the hole and electron capture, respectively)(Table \ref{table_vse}). 
While for \textit{V}$_\mathrm{Se(3)}$, a (2+/+) \ac{TL} of \SI{1.2}{\electronvolt} which is closer to the \ac{CBM} results in a larger energy barrier (1.42 eV) for the hole capture. As a result, the hole capture coefficient is significantly reduced (on the order of \SI{e-15}{\cubic\cm\per\second}) which makes \textit{V}$_\mathrm{Se(3)}$ less detrimental.
Furthermore, even for defect centres with similar order of capture coefficients, their influence on conversion efficiency can differ significantly. 
While \textit{V}$_\mathrm{Se(2)}$ is identified as the most detrimental species in \ce{Sb2Se3}, a much lower (more than two orders of magnitude) concentration of \textit{V}$_\mathrm{Se(1)}$ makes it an inefficient recombination centre.

\newpage

\subsubsection{S5.1.2 \textit{V}$^{2+}_\mathrm{Se(2)}$ $\leftrightarrow$ \textit{V}$^{+}_\mathrm{Se(2)}$ $\leftrightarrow$ \textit{V}$^{0}_\mathrm{Se(2)}$}

\begin{figure}[h!]
    \centering    {\includegraphics[width=0.8\textwidth]{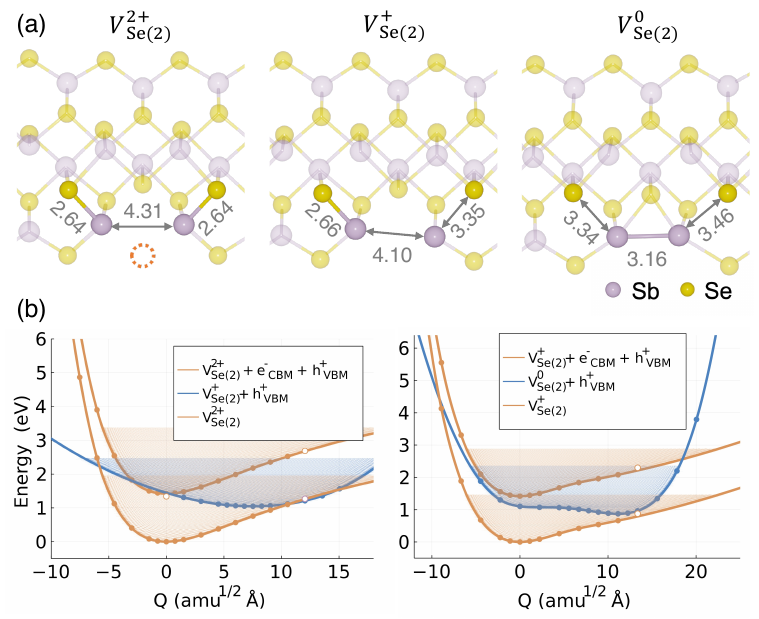}} \\
    \caption{(a) Defect configurations and (b) one-dimensional configuration coordinate diagrams for the charge transitions of \textit{V}$^{2+}_\mathrm{Se}$ $\leftrightarrow$ \textit{V}$^{+}_\mathrm{Se}$ and \textit{V}$^{+}_\mathrm{Se}$ $\leftrightarrow$ \textit{V}$^{0}_\mathrm{Se}$ in \ce{Sb2Se3}. Solid circles are datapoints obtained by DFT calculations and used for fitting, while hollow circles are discarded for fitting due to energy-level crossings. Solid lines represent the best fits to the data.}
    \label{fig_vse2}
\end{figure}

\newpage

The rapid electron and hole capture processes in \textit{V}$^{2+}_\mathrm{Se(2)}$ $\leftrightarrow$ \textit{V}$^{+}_\mathrm{Se(2)}$ largely degrade the performance of \ce{Sb2Se3}. This rapid recombination cycle is ensured, on the one hand, by the rapid hole capture process from \textit{V}$^{2+}_\mathrm{Se(2)}$ to \textit{V}$^{+}_\mathrm{Se(2)}$; and on the other, by slow electron capture from \textit{V}$^{+}_\mathrm{Se(2)}$ to \textit{V}$^{0}_\mathrm{Se(2)}$.
To gain more insights into the capture processes between the charge states of +2, +1 and 0 in \textit{V}$_\mathrm{Se(2)}$, defect configurations and \ac{PESs} of these transitions are compared and shown in Fig. \ref{fig_vse2}.
As we discussed above, the transition between \textit{V}$^{2+}_\mathrm{Se(2)}$ and \textit{V}$^{+}_\mathrm{Se(2)}$ is mainly driven by the shortening/lengthening of one Sb-Se bond around the Se vacancy. While for the transition between \textit{V}$^{+}_\mathrm{Se(2)}$ and \textit{V}$^{0}_\mathrm{Se(2)}$, it is the Sb-Se bond on the other side of the Se vacancy that contributes most to the structural deformation (Fig. \ref{fig_vse2}(a)). As a result, the two neighbouring Sb atoms around the Se vacancy become closer to each other and a Sb-Sb dimer is formed\cite{wang2023four}. The formation of this dimer lowers the total energy of the structure by sharing electrons, and the energy lowering can be seen in Fig. \ref{fig_vse2}(b) from the anharmonicity of the blue curve near \textit{Q}=0.
~\\
~\\
\begin{table*}[ht!]
\caption{Key parameters used to calculate the carrier capture coefficients in the transition of \textit{V}$^{+}_\mathrm{Se(2)}$ $\leftrightarrow$ \textit{V}$^{0}_\mathrm{Se(2)}$: mass-weighted distortion $\Delta$\textit{Q} (amu$^{1/2}$\AA) , energy barrier $\Delta$\textit{E}$_\textrm{b}$  (eV), degeneracy factor \textit{g} of the final state, electron-phonon coupling matrix element $W_{if}$ and scaling factor \textit{s}(\textit{T})\textit{f} at \SI{300}{\kelvin}; and calculated capture coefficient \textit{C} (\SI{}{\cubic\cm\per\second}) and capture cross section $\sigma$ (\SI{}{\cm\squared}) at \SI{300}{\kelvin}}
\label{table_pes}
\begin{tabular*}{\textwidth}{@{\extracolsep{\fill}}c@{\extracolsep{\fill}}c@{\extracolsep{\fill}}c@{\extracolsep{\fill}}c@{\extracolsep{\fill}}c@{\extracolsep{\fill}}c@{\extracolsep{\fill}}c@{\extracolsep{\fill}}c}
    \hline
$\Delta$\textit{Q} & \begin{tabular}[c]{@{}c@{}}Capture\\ process\end{tabular} & $\Delta$\textit{E}$_\textrm{b}$ & \textit{g} & $W_{if}$ & \textit{s}(\textit{T})\textit{f} & \textit{C} & $\sigma$ \\     \hline
\multirow{2}{*}{11.14} & Electron & - & 1 & \SI{4.43e-3}{} &  1.80 &\SI{8.00e-16}{} & \SI{4.05e-23}{} \\
  & Hole & - & 2 & \SI{1.70e-2}{} & - &\SI{1.25e-13}{} & \SI{5.08e-21}{} \\     \hline
\end{tabular*}
\end{table*}

\newpage

\begin{figure}[h!]
    \centering    {\includegraphics[width=0.85\textwidth]{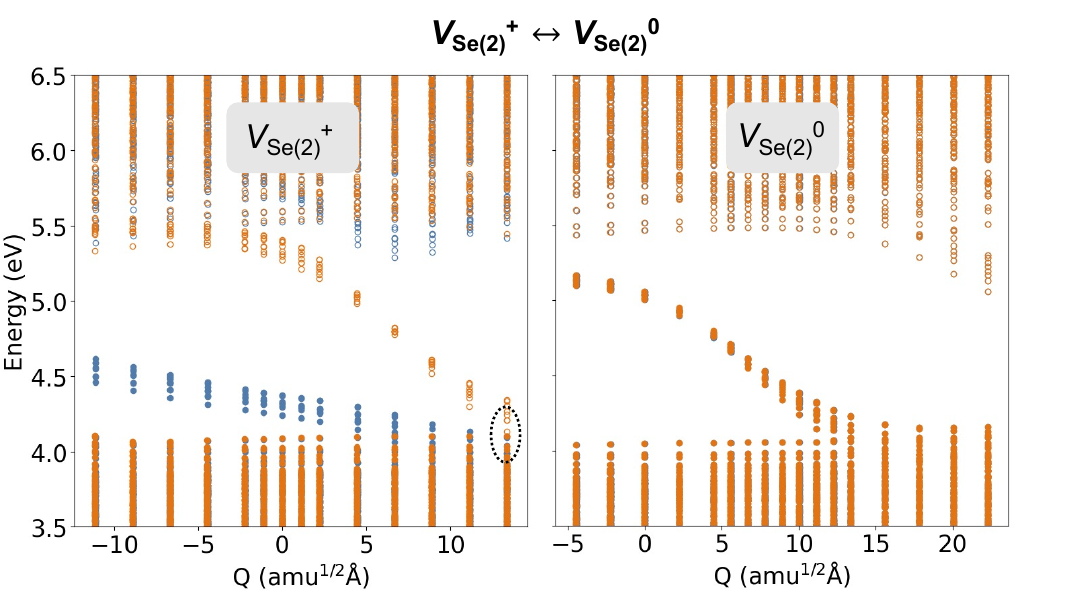}} \\
    \caption{Evolution of Kohn-Sham eigenstates of \textit{V}$\mathrm{_{Se}}^{+}$ and \textit{V}$\mathrm{_{Se}}^{0}$ as a function of the structural deformation \textit{Q}. Solid and hollow circles refer to occupied and unoccupied states, respectively. Different colours represent different spin channels. Dotted ovals represent crossing points which are discarded for fitting.}
    \label{fig_vse2_eig}
\end{figure}
\subsubsection{S5.1.3 \textit{V}$^{+}_\mathrm{Sb}$ $\leftrightarrow$ \textit{V}$^{0}_\mathrm{Sb}$}

\begin{figure}[h!]
    \centering    {\includegraphics[width=0.9\textwidth]{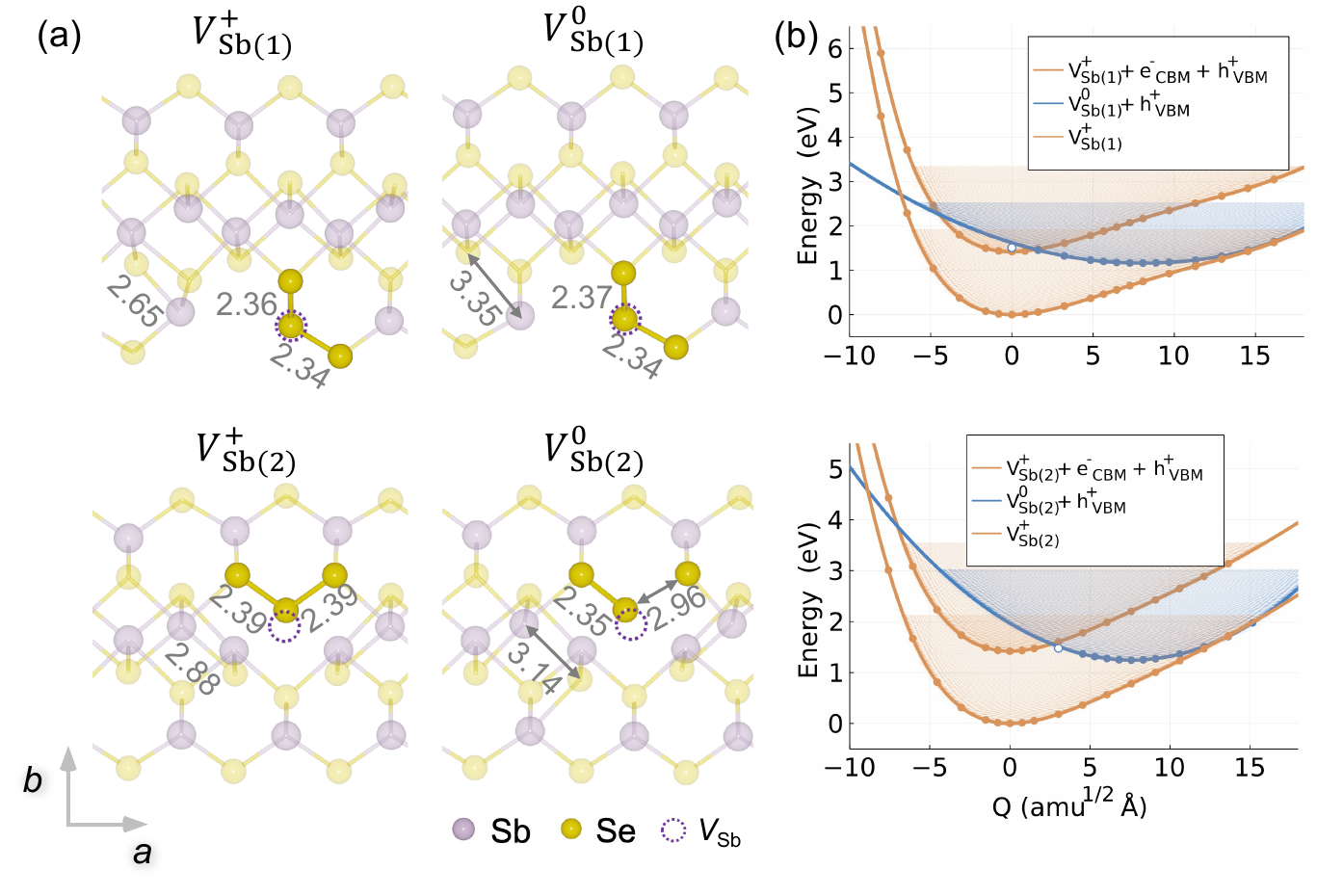}} \\
    \caption{(a) Defect configurations and (b) one-dimensional configuration coordinate diagrams for the charge transitions of \textit{V}$^{+}_\mathrm{Sb}$ $\leftrightarrow$ \textit{V}$^{0}_\mathrm{Sb}$ for both inequivalent sites in \ce{Sb2Se3}. Solid circles are datapoints obtained by DFT calculations and used for fitting, while hollow circles are discarded for fitting due to energy-level crossings. Solid lines represent the best fits to the data.}
    \label{fig_vsb}
\end{figure}

Besides \textit{V}$\mathrm{_{Se(2)}}$, \textit{V}$\mathrm{_{Sb(1)}}$ and \textit{V}$\mathrm{_{Sb(2)}}$ also lead to a $V_\mathrm{OC}$ loss under Se-rich conditions, which results from the rapid recombination cycle of \textit{V}$^{+}_\mathrm{Sb}$ $\leftrightarrow$ \textit{V}$^{0}_\mathrm{Sb}$.
Defect configurations and \ac{PESs} of this transition are shown in Fig. \ref{fig_vsb}.
When \textit{V}$_\mathrm{Sb}$ is at +1 charge state, a Se-Se-Se trimer is formed around the Sb vacancy to pair electrons and lower the energy \cite{wang2023four}. This can also be viewed as a defect complex of \textit{V}$\mathrm{_{Se}}$ and Se$\mathrm{_{Sb}}$.
Similar to \textit{V}$^{2+}_\mathrm{Se}$ $\leftrightarrow$ \textit{V}$^{+}_\mathrm{Se}$, the structural deformation during the transition of \textit{V}$^{+}_\mathrm{Sb}$ $\leftrightarrow$ \textit{V}$^{0}_\mathrm{Sb}$ is related to the change of one Sb-Se bond near the Sb vacancy (Fig. \ref{fig_vsb}). This further results in similar shapes of \ac{PESs} to \textit{V}$^{2+}_\mathrm{Se}$ $\leftrightarrow$ \textit{V}$^{+}_\mathrm{Se}$, which contribute to reasonable capture coefficients (Table \ref{table_vsb}).

~\\
~\\
~\\
\begin{table*}[h!]
\caption{Key parameters used to calculate the carrier capture coefficients in the transition of \textit{V}$^{+}_\mathrm{Sb}$ $\leftrightarrow$ \textit{V}$^{0}_\mathrm{Sb}$ for different inequivalent sites: mass-weighted distortion $\Delta$\textit{Q} (amu$^{1/2}$\AA), energy barrier $\Delta$\textit{E}$_\textrm{b}$  (meV), degeneracy factor \textit{g} of the final state, electron-phonon coupling matrix element $W_{if}$ and scaling factor \textit{s}(\textit{T})\textit{f} at \SI{300}{\kelvin}; and calculated capture coefficient \textit{C} (\SI{}{\cubic\cm\per\second}) and capture cross section $\sigma$  (\SI{}{\cm\squared}) at \SI{300}{\kelvin}}
\label{table_vsb}
\begin{tabular*}{\textwidth}{@{\extracolsep{\fill}}c@{\extracolsep{\fill}}c@{\extracolsep{\fill}}c@{\extracolsep{\fill}}c@{\extracolsep{\fill}}c@{\extracolsep{\fill}}c@{\extracolsep{\fill}}c@{\extracolsep{\fill}}c@{\extracolsep{\fill}}c}
    \hline
Species & $\Delta$\textit{Q} & \begin{tabular}[c]{@{}c@{}}Capture\\ process\end{tabular} & $\Delta$\textit{E}$_\textrm{b}$ & \textit{g} & $W_{if}$ & \textit{s}(\textit{T})\textit{f} & \textit{C} & $\sigma$ \\     \hline
\multirow{2}{*}{\textit{V}$_\mathrm{Sb(1)}$} &\multirow{2}{*}{8.07} & Electron & \SI{48}{} & 2 & \SI{1.01e-2}{} & 1.80 &\SI{5.88e-8}{} & \SI{2.98e-15}{} \\
 & & Hole & \SI{1538}{} & 1 & \SI{3.50e-2}{} & - &\SI{8.77e-11}{} & \SI{7.12e-18}{} \\     \hline
\multirow{2}{*}{\textit{V}$_\mathrm{Sb(2)}$} &\multirow{2}{*}{7.57} & Electron & \SI{140}{} & 2  & \SI{1.95e-2}{} & 1.80  &\SI{1.84e-9}{} & \SI{1.40e-17}{} \\
 & & Hole & \SI{382}{} & 1 & \SI{3.65e-2}{} & - &\SI{7.88e-11}{} & \SI{3.36e-19}{} \\     \hline
\end{tabular*}
\end{table*}
~\\
~\\
\begin{figure}[h!]
    \centering    {\includegraphics[width=0.8\textwidth]{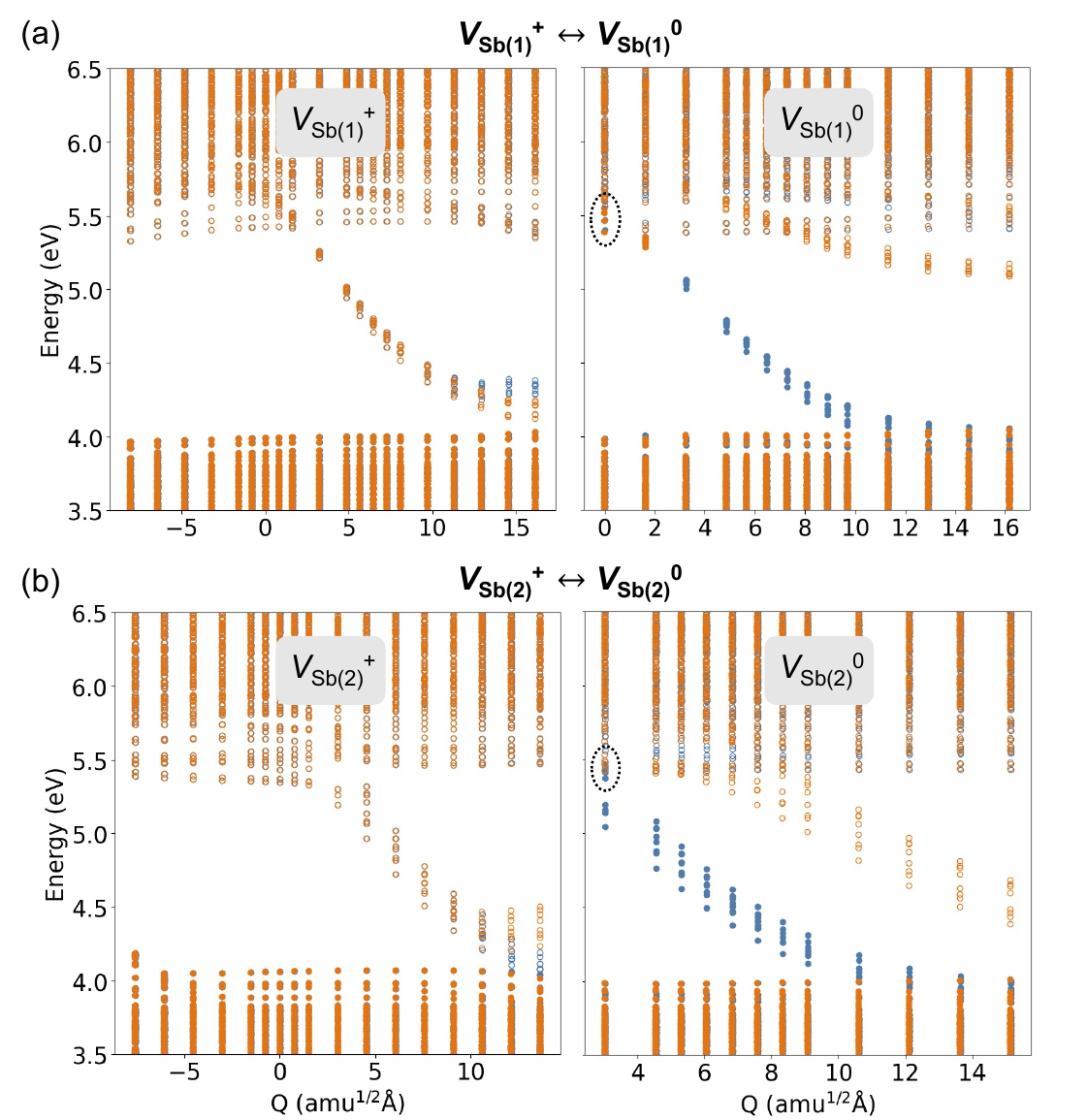}} \\
    \caption{Evolution of Kohn-Sham eigenstates of \textit{V}$\mathrm{_{Sb}}^{+}$ and \textit{V}$\mathrm{_{Sb}}^{0}$ as a function of the structural deformation \textit{Q} for both inequivalent sites. Solid and hollow circles refer to occupied and unoccupied states, respectively. Different colours represent different spin channels. Dotted ovals represent crossing points which are discarded for fitting.}
    \label{fig_vsb_eig}
\end{figure}

\newpage

\subsection{S5.2 Antisites}
\begin{figure}[h!]
    \centering    {\includegraphics[width=\textwidth]{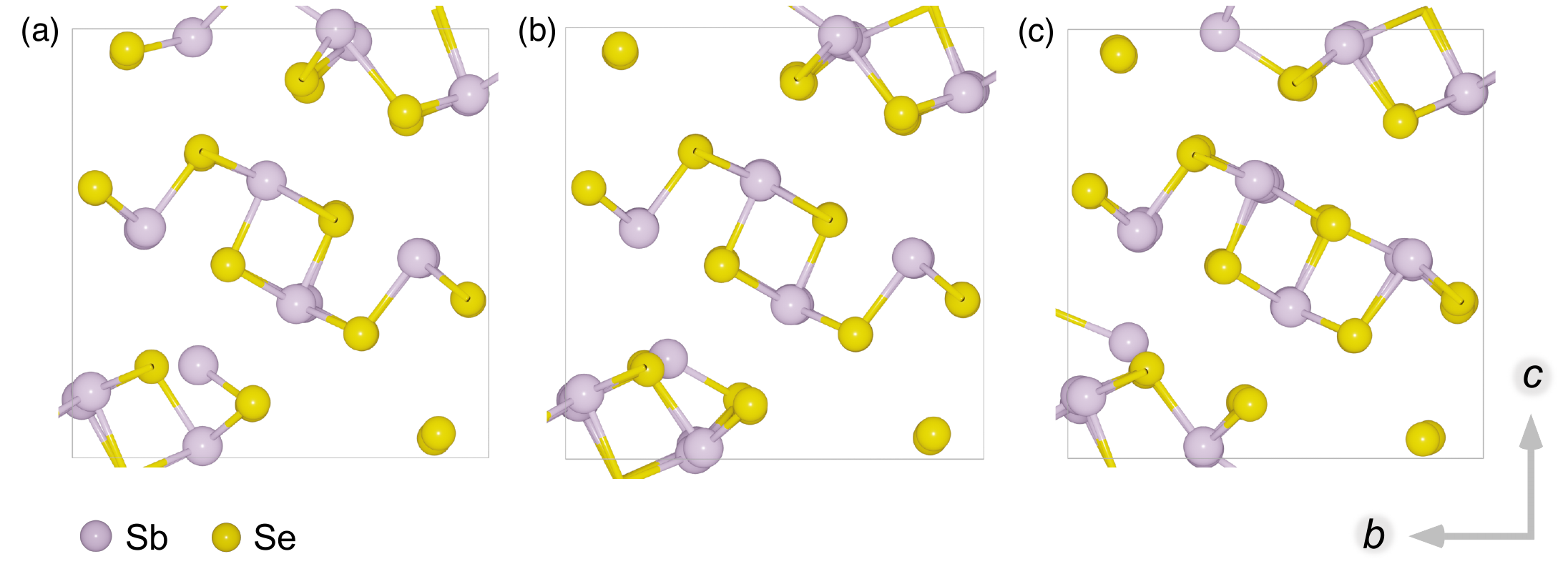}} \\
    \caption{Defect configurations of (a) Sb$\mathrm{_{Se(3)}}^{3+}$, (b) metastable Sb$\mathrm{_{Se(3)}}^{2+ *}$ and (c) Sb$\mathrm{_{Se(3)}}^{2+}$.}
    \label{fig_stru}
\end{figure}

\begin{figure}[h!]
    \centering    {\includegraphics[width=0.5\textwidth]{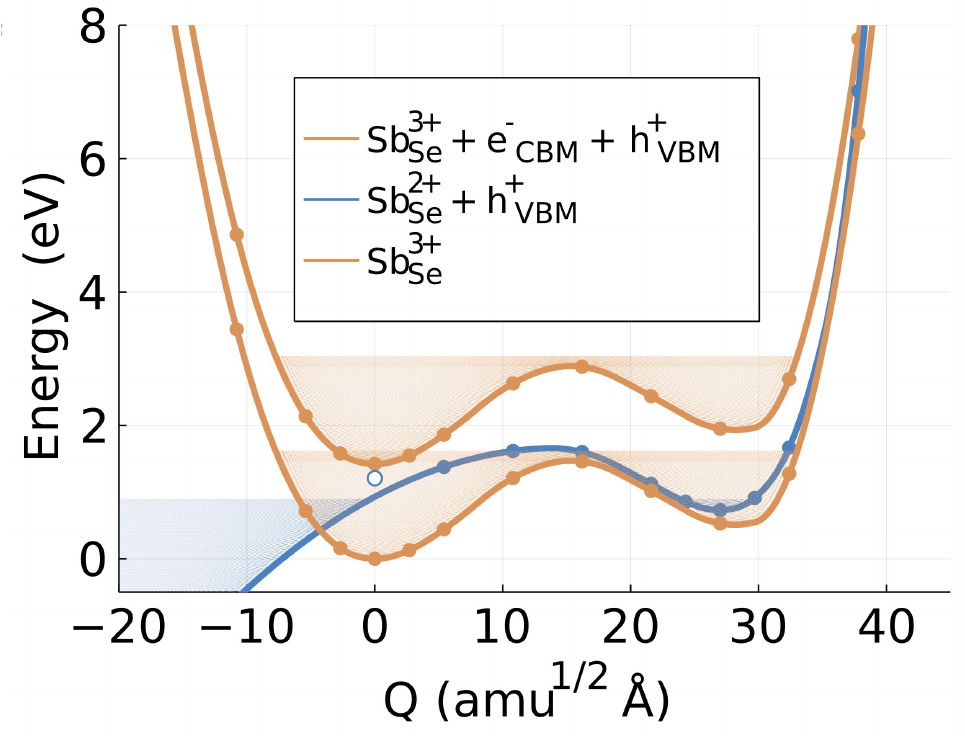}} \\
    \caption{One-dimensional configuration coordinate diagrams for the charge transition between Sb$\mathrm{_{Se(3)}}^{3+}$ and Sb$\mathrm{_{Se(3)}}^{2+}$ in \ce{Sb2Se3}. Solid circles are datapoints obtained by DFT calculations and used for fitting, while hollow circles are discarded for fitting due to energy-level crossings. Solid lines represent the best fits to the data.}
    \label{fig_ant_cc}
\end{figure}

\begin{figure}[h!]
    \centering    {\includegraphics[width=\textwidth]{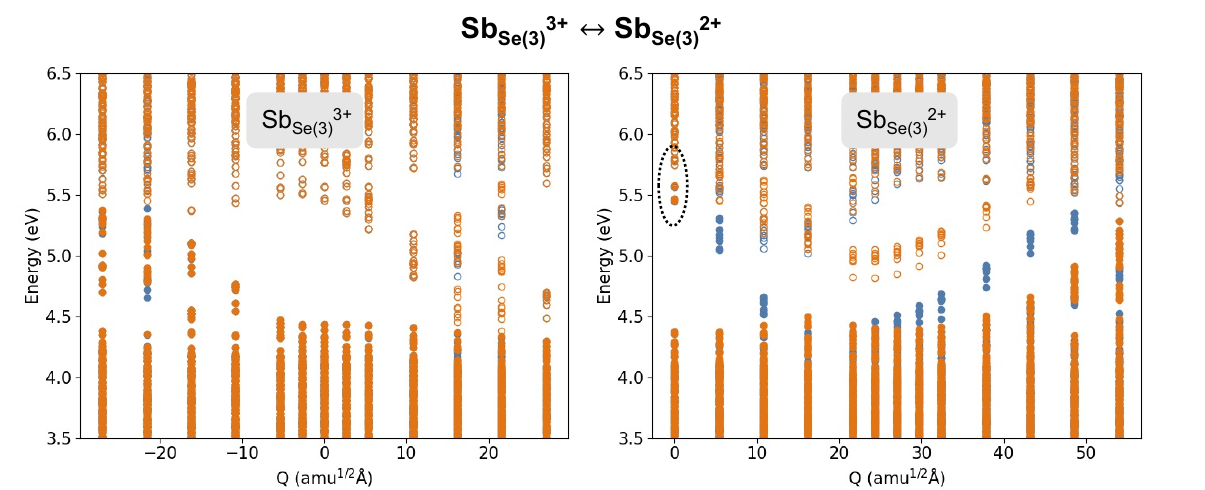}} \\
    \caption{Evolution of Kohn-Sham eigenstates of Sb$\mathrm{_{Se(3)}}^{3+}$ and Sb$\mathrm{_{Se(3)}}^{2+}$ as a function of the structural deformation \textit{Q}. Solid and hollow circles refer to occupied and unoccupied states, respectively. Different colours represent different spin channels. Dotted ovals represent crossing points which are discarded for fitting.}
    \label{fig_ant_eig}
\end{figure}

If two defect configurations exhibit substantial dissimilarity from each other with large $\Delta$\textit{Q}, we assume it would be difficult for the carrier capture to occur directly. In these cases, metastable defect configurations should be considered as intermediate configurations\cite{kavanagh2022impact} (with superscript asterisks in the figure). 
Here we take the transition between Sb$\mathrm{_{Se(3)}}^{3+}$ and Sb$\mathrm{_{Se(3)}}^{2+}$ as an example. 
\ac{PESs} show double-well shapes in the \ac{CC} diagram (Fig. \ref{fig_ant_cc}), which suggest metastable defect configurations.
The electronic eigenstates as a function of \textit{Q} were carefully checked and datapoints with defect levels crossing the band edges are not used for fitting (Fig. \ref{fig_ant_eig}).
A huge $\Delta$\textit{Q} of 27 amu$^{1/2}${\AA } for this transition indicates a large displacement of the Sb$\mathrm{_{Se(3)}}$ defect when it changes the charge state from +3 to +2 (as shown in Fig. \ref{fig_stru}(a) and (c)). Thus, a metastable Sb$\mathrm{_{Se(3)}}^{2+ *}$ (Fig. \ref{fig_stru}(b)) is further calculated by taking the configuration of Sb$\mathrm{_{Se(3)}}$ as an initial structure for relaxation with +2 charge state.
~\\
~\\
~\\
~\\
~\\
~\\

\newpage

\begin{figure}[h!]
    \centering    {\includegraphics[width=0.6\textwidth]{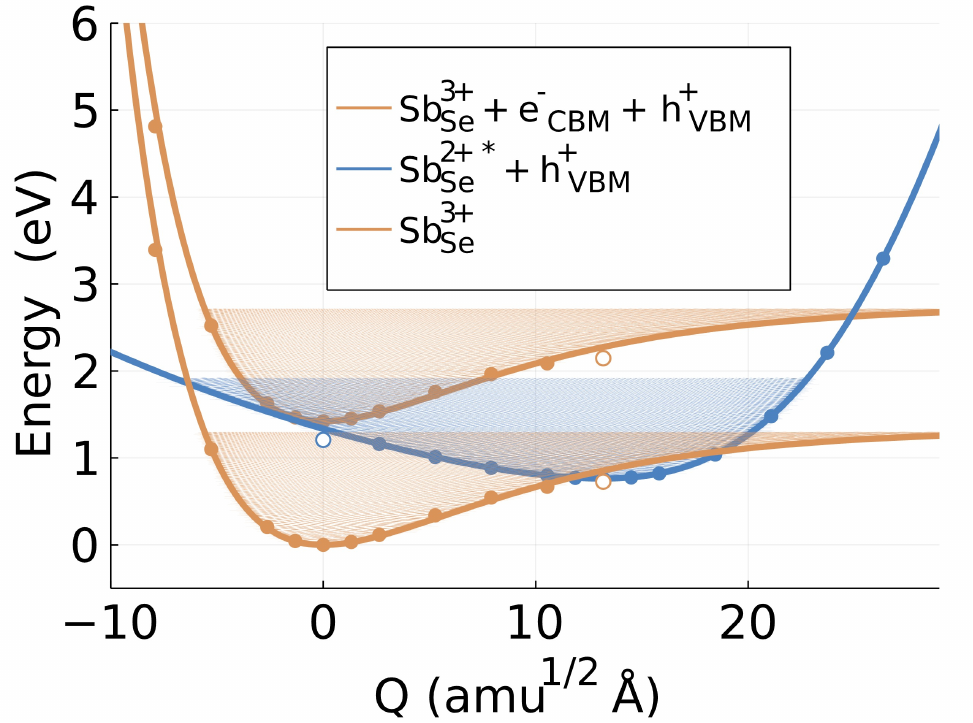}} \\
    \caption{One-dimensional configuration coordinate diagrams for the charge transition between Sb$\mathrm{_{Se(3)}}^{3+}$ and metastable Sb$\mathrm{_{Se(3)}}^{2+ *}$ in \ce{Sb2Se3}. Solid circles are datapoints obtained by DFT calculations and used for fitting, while hollow circles are discarded for fitting due to energy-level crossings. Solid lines represent the best fits to the data.}
    \label{fig_ant_cc_meta}
\end{figure}

\begin{table*}[ht!]
\caption{Key parameters used to calculate the carrier capture coefficients in the transition of Sb$\mathrm{_{Se(3)}}^{3+}$ $\leftrightarrow$ Sb$\mathrm{_{Se(3)}}^{2+ *}$: mass-weighted distortion $\Delta$\textit{Q} (amu$^{1/2}$\AA) , energy barrier $\Delta$\textit{E}$_\textrm{b}$  (meV), degeneracy factor \textit{g} of the final state, electron-phonon coupling matrix element $W_{if}$ and scaling factor \textit{s}(\textit{T})\textit{f} at \SI{300}{\kelvin}; and calculated capture coefficient \textit{C} (\SI{}{\cubic\cm\per\second}) and capture cross section $\sigma$ (\SI{}{\cm\squared}) at \SI{300}{\kelvin}}
\label{table_pes}
\begin{tabular*}{\textwidth}{@{\extracolsep{\fill}}c@{\extracolsep{\fill}}c@{\extracolsep{\fill}}c@{\extracolsep{\fill}}c@{\extracolsep{\fill}}c@{\extracolsep{\fill}}c@{\extracolsep{\fill}}c@{\extracolsep{\fill}}c}
    \hline
$\Delta$\textit{Q} & \begin{tabular}[c]{@{}c@{}}Capture\\ process\end{tabular} & $\Delta$\textit{E}$_\textrm{b}$ & \textit{g} & $W_{if}$ & \textit{s}(\textit{T})\textit{f} & \textit{C} & $\sigma$ \\     \hline
\multirow{2}{*}{13.18} & Electron &1206  & 2 & \SI{4.30e-2}{} & 1.39  &\SI{1.39e-6}{} & \SI{7.04e-14}{} \\
  & Hole & \SI{315}{} & 1 & \SI{7.86e-3}{} & 0.51 &\SI{3.10e-8}{} & \SI{2.52e-15}{} \\     \hline
\end{tabular*}
\end{table*}

\begin{figure}[h!]
    \centering    {\includegraphics[width=\textwidth]{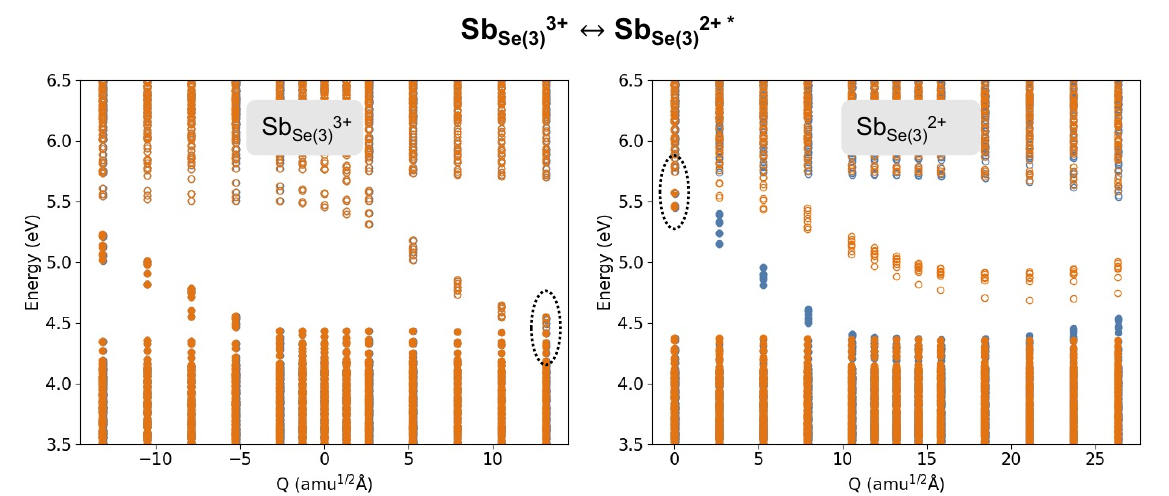}} \\
    \caption{Evolution of Kohn-Sham eigenstates of Sb$\mathrm{_{Se(3)}}^{3+}$ and metastable Sb$\mathrm{_{Se(3)}}^{2+ *}$ as a function of the structural deformation \textit{Q}. Solid and hollow circles refer to occupied and unoccupied states, respectively. Different colours represent different spin channels. Dotted ovals represent crossing points which are discarded for fitting.}
    \label{fig_ant_eig_meta}
\end{figure}

\begin{figure}[h!]
    \centering    {\includegraphics[width=0.55\textwidth]{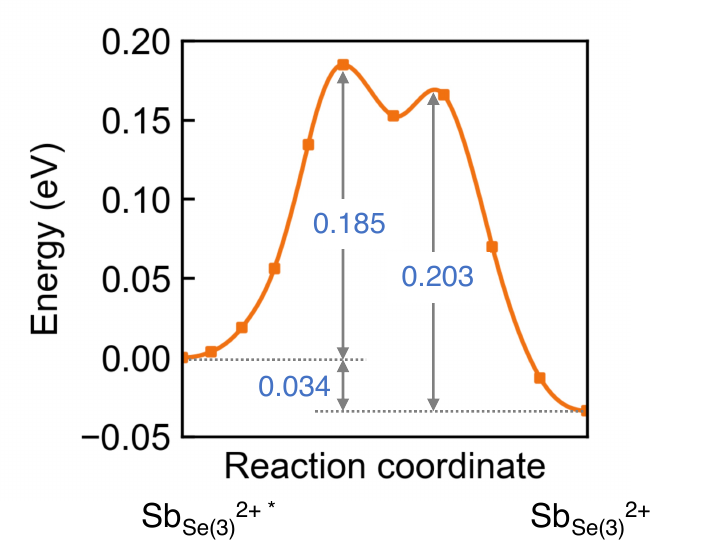}} \\
    \caption{Potential energy surface for the transformation between Sb$\mathrm{_{Se(3)}}^{2+}$ and Sb$\mathrm{_{Se(3)}}^{3+}$.}
    \label{fig_neb}
\end{figure}

\begin{figure}[h!]
    \centering    {\includegraphics[width=0.65\textwidth]{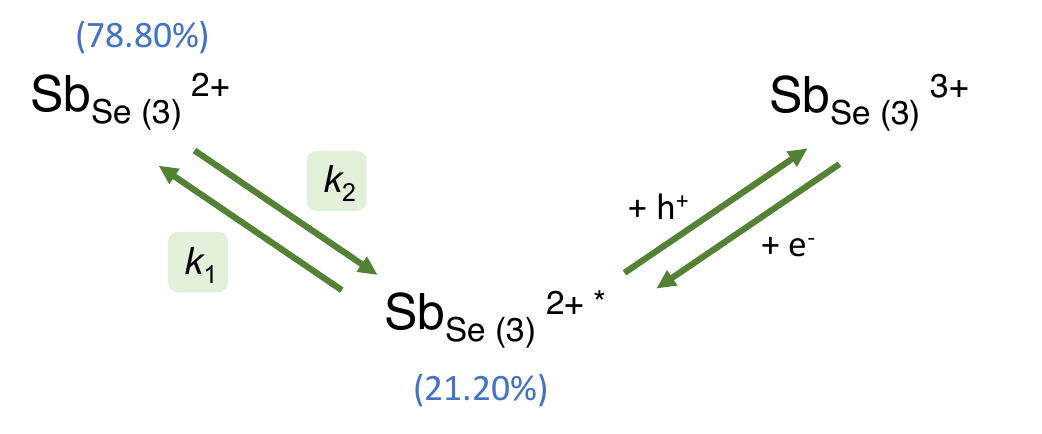}} \\
    \caption{Schematic diagram of the transition between Sb$\mathrm{_{Se(3)}}^{2+}$,  Sb$\mathrm{_{Se(3)}}^{2+ *}$ and Sb$\mathrm{_{Se(3)}}^{3+}$.}
    \label{fig_dia}
\end{figure}

The transition pathway is calculated by climbing image-nudged elastic band (CI-NEB) implemented within VASP (as shown in Fig. \ref{fig_neb}). 
Transition State Theory\cite{vineyard1957frequency} is used to calculate the rate \textit{k} of transition between Sb$\mathrm{_{Se(3)}}^{2+}$ and Sb$\mathrm{_{Se(3)}}^{2+ *}$,
\begin{equation}
k=\nu g \mathrm{exp}(-\frac{\Delta E}{k_BT})
\end{equation}
where $\nu$ is the attempt frequency which is calculated from the \ac{PESs} between configurations, \textit{g} is the ratio between the degeneracies of the initial and final states and $\Delta E$ is the energy barrier.

Let's refer to the transition from metastable Sb$\mathrm{_{Se(3)}}^{2+ *}$ to Sb$\mathrm{_{Se(3)}}^{2+}$ as process 1 and from Sb$\mathrm{_{Se(3)}}^{2+}$ to Sb$\mathrm{_{Se(3)}}^{2+ *}$ as process 2 (as shown in the schematic diagram in Fig. \ref{fig_dia}).
Activation energy barriers of $\Delta E_1$=\SI{185}{\meV} and $\Delta E_2$=\SI{203}{\meV} and frequencies of $\nu_1$=\SI{0.95}{\THz} and $\nu_2$=\SI{1.84}{\THz} are obtained from the \ac{PESs}. These give transition rate $k_1$ = \SI{7.52e8}{\per\second} and $k_2$= \SI{7.26e8}{\per\second} at room temperature, indicating fast rates for both transitions and thus unlikely to be major bottlenecks in a recombination cycle.

According to Boltzmann distribution,
\begin{equation}
\frac{p_{\mathrm{Sb{_{Se(3)}}^{2+ *}}}}{p_{\mathrm{Sb{_{Se(3)}}^{2+}}}}=e^{-\frac{E_{\mathrm{Sb{_{Se(3)}}^{2+ *}}}-E_{\mathrm{Sb{_{Se(3)}}^{2+}}}}{k_BT}}
\end{equation}
where $E_\mathrm{Sb{_{Se(3)}}^{2+ *}}$ and $E_\mathrm{Sb{_{Se(3)}}^{2+}}$ are total energies of Sb$\mathrm{_{Se(3)}}^{2+ *}$ to Sb$\mathrm{_{Se(3)}}^{2+}$.
So we get $p_{\mathrm{Sb{_{Se(3)}}^{2+ *}}}$=21.20\% and $p_{\mathrm{Sb{_{Se(3)}}^{2+}}}$=78.80\%. This means when $\mathrm{Sb{_{Se(3)}}^{2+ *}}$ and $\mathrm{Sb{_{Se(3)}}^{2+}}$ are in equilibrium, the relative populations are 21.20\% of $\mathrm{Sb{_{Se(3)}}^{2+ *}}$ and 78.80\% of $\mathrm{Sb{_{Se(3)}}^{2+}}$.

Despite large capture coefficients for both electron and hole capture processes in the transition of $\mathrm{Sb{_{Se(3)}}^{3+}}$ $\leftrightarrow$ $\mathrm{Sb{_{Se(3)}}^{2+ *}}$, $\mathrm{Sb{_{Se(3)}}}$ is predicted to be benign with negligible impact on the performance. 
This is because the thermodynamically stable charge state for $\mathrm{Sb{_{Se(3)}}}$ under all chemical potentials and growth temperatures (without extrinsic doping) is -1, and the $\mathrm{Sb{_{Se(3)}}}$ hole capture rates from -1 to +3 are calculated to be extremely slow, resulting in negligible Sb$\mathrm{_{Se(3)}}^{3+}$ concentrations both in the dark and under illumination.


\newpage

\section{S6. Anisotropic optical absorption spectra}

\begin{figure}[h!]
    \centering    {\includegraphics[width=0.4\textwidth]{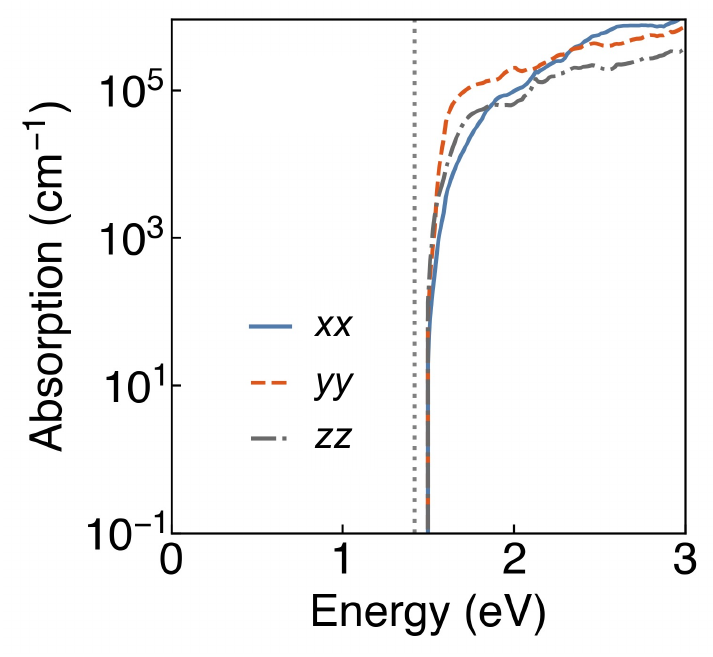}} \\
    \caption{Calculated optical absorption spectra of \ce{Sb2Se3} arising from direct valence to conduction band transitions. \textit{xx}, \textit{yy} and \textit{zz} represent the directions of the electric polarisation vector of light. The dotted line indicates the calculated fundamental (indirect) band gap.}
    \label{fig_abs}
\end{figure}

\section{S7. Equilibrium defect concentration and trap-limited conversion efficiency at 648 K}

\begin{figure}[h!]
    \centering    {\includegraphics[width=0.8\textwidth]{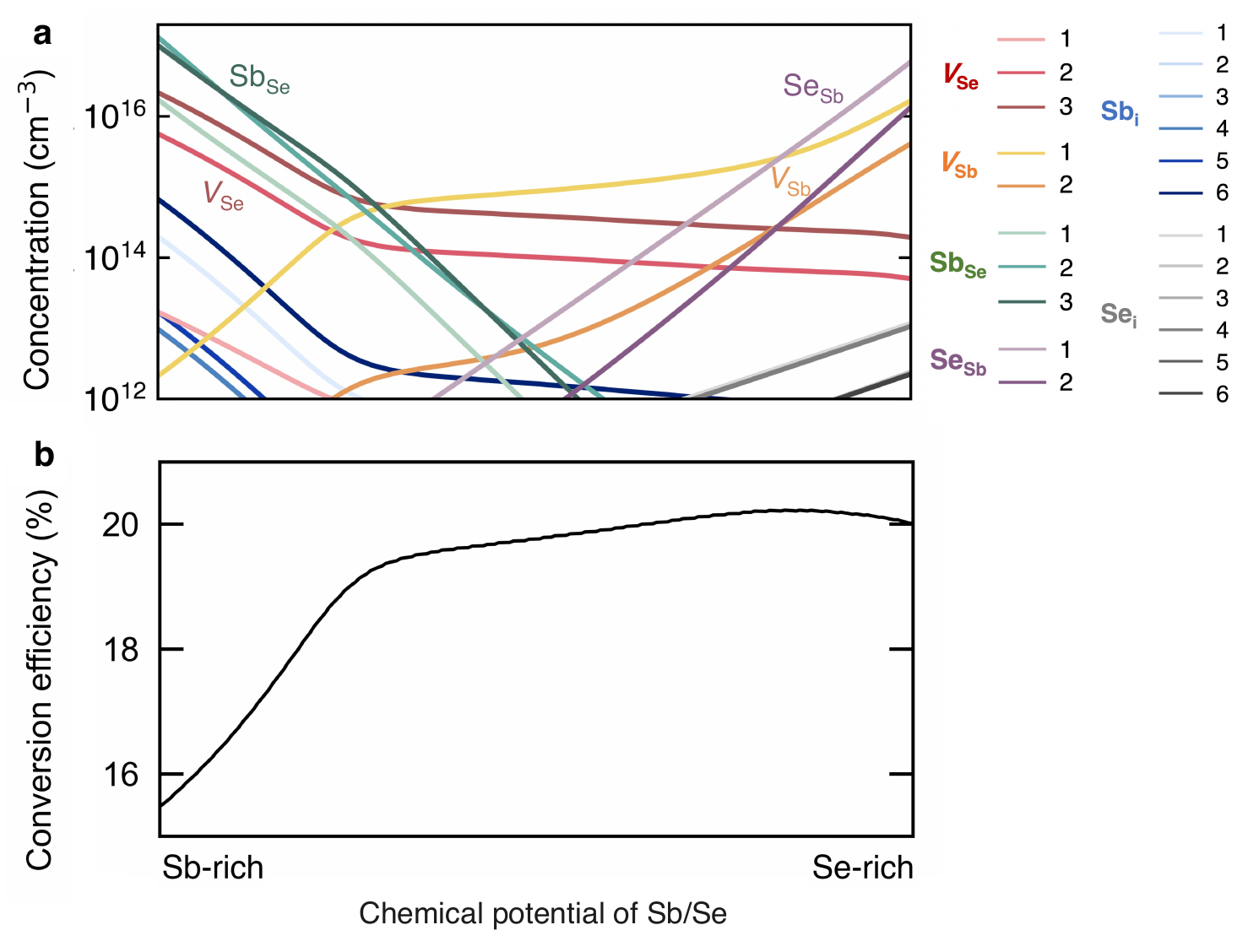}} \\
    \caption{(a) Equilibrium defect concentration and (b) trap-limited conversion efficiency as a function of the growth condition at \SI{300}{\kelvin} in \ce{Sb2Se3} crystals grown at \SI{648}{\kelvin}.}
    \label{fig_375C}
\end{figure}

To understand the effect of annealing temperature on the performance of \ce{Sb2Se3}, equilibrium defect concentration at T$_\textrm{anneal}$ = \SI{648}{\kelvin} was also calculated, which matches the value used in the champion \ce{Sb2Se3} devices\cite{zhao2022regulating}. 
As shown in Fig. \ref{fig_375C}(a), defect concentrations are two orders of magnitude higher than those annealed at \SI{550}{\kelvin}. Higher vacancy concentrations lead to lower conversion efficiency in \ce{Sb2Se3}, with a range of \SI{15.5}{\percent} to \SI{20.2}{\percent} depending on the growth conditions (Fig. \ref{fig_375C}(b)).

\newpage
\bibliography{references}